\documentclass[iop, tighten]{emulateapj}

\usepackage{pslatex}
\usepackage{amsmath,amssymb}
\usepackage{textcomp}

\usepackage[]{hyperref}
\hypersetup {
    bookmarks=true,                    
    pdftitle={Magnetic Inhibition of Convection and the Fundamental Properties of Low-Mass Stars. II. Fully Convective Main Sequence Stars},                   
    pdfauthor={Gregory A. Feiden},     
    pdfsubject={},                     
    colorlinks=true,                   
    linkcolor=blue,                    
    citecolor=blue,                    
    urlcolor=blue                     
}
\newcommand{\myemail}{gregory.a.feiden@gmail.com}
\newcommand{\bcemail}{brian.chaboyer@dartmouth.edu}

\newcommand{\msun}{M_{\odot}}                
\newcommand{\rsun}{R_{\odot}}                
\newcommand{\teff}{T_{\rm eff}}              
\newcommand{\amlt}{\alpha_{\rm MLT}}         
\newcommand{\dela}{\nabla_{\rm ad}}          
\newcommand{\delx}{\nabla_{\rm \chi}}        
\newcommand{\tgrad}{\nabla_{\rm s}}          
\newcommand{\uconv}{u_{\rm conv}}            
\newcommand{\deltamm}{\delta_{\rm MM}}       

\defcitealias{Doyle2011}{D11}
\defcitealias{Bender2012}{B12}

\shortauthors{Feiden \& Chaboyer}
\slugcomment{Accepted to the Astrophysical Journal; \today}

\begin{document}

\title{Magnetic Inhibition of Convection and the Fundamental Properties \\
of Low-Mass Stars. II. Fully Convective Main Sequence Stars}

\author{Gregory A. Feiden\altaffilmark{1} and Brian Chaboyer\altaffilmark{2}}
\affil{$^1$ Department of Physics and Astronomy, Uppsala University, Box 516, 
       SE-751 20 Uppsala, Sweden; \href{mailto:\myemail}{\myemail} \\
       $^2$ Department of Physics and Astronomy, Dartmouth College, 6127 Wilder 
       Laboratory, Hanover, NH 03755, USA;
\href{mailto:\bcemail}{\bcemail}}

\begin{abstract}
We examine the hypothesis that magnetic fields are inflating the radii of
fully convective main sequence stars in detached eclipsing binaries (DEBs). The
magnetic Dartmouth stellar evolution code is used to analyze two
systems in particular: Kepler-16 and CM Draconis. Magneto-convection is treated
assuming stabilization of convection and also by assuming reductions in convective
efficiency due to a turbulent dynamo. We find that magnetic stellar models
are unable to reproduce the properties of inflated fully convective main sequence
stars, unless strong interior magnetic fields in excess of 10~MG are present. 
Validation of the magnetic field hypothesis given the current generation of
magnetic stellar evolution models therefore depends critically on
whether the generation and maintenance of strong interior magnetic fields is
physically possible. An examination of this requirement is provided.
Additionally, an analysis of previous studies invoking the influence of star
spots is presented to assess the suggestion that star spots are inflating 
stars and biasing light curve analyses toward larger radii.
From our analysis, we find that there is not yet sufficient evidence to 
definitively support the hypothesis that magnetic fields are responsible 
for the observed inflation among fully convective main sequence stars in 
DEBs. 
\end{abstract}
\keywords{binaries: eclipsing -- stars: evolution -- stars: interiors
-- stars: low-mass -- stars: magnetic field}

\section{Introduction}
\label{sec:intro}

Outer layers of low-mass stars are unstable to thermal convection due to a rapid
increase in opacity resulting from the partial ionization of hydrogen and the 
dissociation of H$_2$. Below $0.35\msun$, main sequence 
stellar interiors are theorized to become fully convective along the main sequence
\citep{Limber1958a,CB97}. Largely characterized by near-adiabatic convection,
fully convective stars are considered the simplest stars to describe from
a theoretical perspective. In fact, properties of fully convective stars
predicted by stellar structure models are largely insensitive to input
variables (e.g., the mixing length parameter, $\amlt$) and input physics \citep[e.g., nuclear 
reaction rates, element diffusion;][]{CB97,Dotter2007}. Discovery
of significant radius discrepancies between observations and stellar model
predictions for fully convective stars therefore presents a curious puzzle 
\citep[see, e.g.,][]{Torres2010,FC12}.

Evidence indicating that stellar structure models can not properly predict
radii of fully convective stars has been gathered from studies of detached
eclipsing binaries (DEBs). Masses and radii can be measured for stars in 
DEBs with precisions below 3\% provided the observations are of high quality 
and analyses are performed with
care \citep{Popper1984,Andersen1991,Torres2010}. Presently, there are five
DEBs with at least one
fully convective component whose mass and radius has been quoted with
precision below 3\%: Kepler-38 \citep{Kep38}, Kepler-16 \citep{Doyle2011,
Winn2011,Bender2012}, LSPM J1112+7626 \citep{Irwin2011}, KOI-126 
\citep{Carter2011}, and CM Draconis \citep[hereafter CM Dra;][]{Lacy1977,Metcalfe1996,
Morales2009a}. Of these systems, only the fully convective stars of KOI-126 
can be accurately characterized by stellar evolution models \citep{Feiden2011,
SD12}. Every other fully convective star appears to have a radius inflated
compared to model predictions.

Most consequential are the inflated radii of the stars in CM Dra. Historically,
the stars of CM Dra are \emph{the} fully convective stars against which
to benchmark stellar models. As such, CM Dra has been well-studied and 
rigorously characterized. Over the years, discrepancies between model
radii of CM Dra and those determined from observations has grown. Initial
modeling efforts were optimistic that agreement could be achieved \citep{CB1995}, 
but disagreement was quickly identified
with the introduction of more sophisticated models \citep{BCAH98} and 
more precise mass and radius measurements \citep{Metcalfe1996,Morales2009a}.
This disparity
has been increased, yet again, with converging reports of the system's 
metallicity \citep{RojasAyala2012,Terrien2012}.

Strong magnetic fields maintained by tidal synchronization are presently 
considered the leading culprit producing the observed radius discrepancies 
\citep[e.g.,][]{MM01,Ribas2006,Lopezm2007,Chabrier2007,Morales2008,Morales2009a,MM11}. Magnetic
activity indicators, such as soft X-ray emission, Ca {\sc ii} H \& K emission,
and H$\alpha$ emission, appear to correlate with radius inflation \citep{
Lopezm2007,FC12,Stassun2012}, providing evidence in favor of the magnetic
hypothesis. Theoretical investigations also support a magnetic origin of
radius inflation for main sequence DEB stars \citep{Chabrier2007,Morales2010,MM11,
FC13}.

Despite significant evidence in favor of the magnetic hypothesis, several
clues suggest otherwise. Discovery of the hierarchical triple KOI-126 in 2011 introduced a second
pair of well-characterized fully convective stars whose masses and radii
were measured with better than 2\% precision. Stellar evolution models 
are able to reproduce the properties of KOI-126 (B,\,C),
as previously mentioned, based only on inferred properties from the more massive
primary star \citep{Feiden2011,SD12}. With orbital and stellar properties similar 
to CM Dra, KOI-126 
(B,\,C) presents a sharp contrast to known modeling disagreements.

Adding to the evidence mounting against the magnetic hypothesis, fully convective 
stars in Kepler-16, Kepler-38, and LSPM-J1112+7626 show inflated radii despite 
existing in long period ($>$ 17\,days) systems. The fact that most inflated
stars in DEBs appeared to exist in short period systems was proffered as
strong circumstantial evidence in support of the magnetic hypothesis. However,
only a few DEB systems were known, all of which had short orbital periods
due to inherent observational biases. With the influx of data from long time 
baseline photometric monitoring campaigns, including \emph{Kepler} and MEarth,
fully convective stars in long period DEBs have been shown to have inflated radii. 

Of course, the presence of inflated low-mass stars in long period systems is 
only contradictory if the inflated stars are slowly rotating 
($v\sin i \lesssim 5$\,km\,s$^{-1}$). \citet{Irwin2011} find that LSPM~J1112+7626~A 
rotates with a period of 65\,days, from which we infer an age of order 9\,Gyr
assuming the gyrochronology relation of \citet{Barnes2010}. This result
requires confirmation, but if confirmed, it would be seem likely that the 
fully convective, low-mass secondary is slowly rotating. A rotation period has also been 
measured for the primary star in Kepler-16 \citep{Winn2011}. It was determined
to be rotating with a period of nearly 36\,days, close to the pseudo-synchronization
rotation period \citep{Hut1981}. If we assume the fully convective secondary 
has a similar rotation period (from pseudo-synchronization), then it would 
have $v\sin i < 0.5$ km\,s$^{-1}$, well below the empirical velocity threshold 
thought to be required for fully convective stars to maintain a strong magnetic
field \citep{Reiners2009a}. Therefore, in at least two cases, it appears that
slowly rotating fully convective stars exhibit inflated radii.

In this paper, we extend our on-going investigation into the magnetic origin 
of inflated low-mass stellar radii, initiated in \citet{FC12b,FC13},
to fully convective stars. A brief overview of how we include magnetic effects
in our models is presented in Section \ref{sec:dmestar}. Detailed analysis
of two well-characterized DEBs, Kepler-16 and CM Dra, is given in Section
\ref{sec:ind_deb} with a discussion of the results in Section \ref{sec:disc}.
Section \ref{sec:disc} also provides comparisons with previous studies 
and a careful examination of the magnetic hypothesis. A summary of 
key results and conclusions is then given in Section \ref{sec:summ}.

\section{Magnetic Dartmouth Stellar Evolution Code}
\label{sec:dmestar}

Stellar evolution models used in this investigation are from the Dartmouth
Magnetic Evolutionary Stellar Tracks and Relations (DMESTAR) program. DMESTAR 
was developed as an extension of the Dartmouth Stellar Evolution Program 
\citep[DSEP;][]{Dotter2008}, a descendant of the Yale Rotating Evolution 
Code \citep{Guenther1992}. The magnetic version of the Dartmouth stellar
evolution code is described in \citet{FC12b}, \citet{FeidenTh}, and \citet{FC13}.
We refer the reader to these papers for a thorough overview.

\subsection{Physics for Fully Convective Models}

The pertinent aspects of the standard, non-magnetic stellar evolution code 
for modeling fully convective stars are: the equation of state (EOS) and
the surface boundary conditions. All fully convective stars are modeled 
with the FreeEOS, a publicly available EOS code written by Alan Irwin and
based on the free energy minimization technique.\footnote{Available at 
http://www.freeeos.sourceforge.net/}
We call FreeEOS in the EOS4 configuration to provide a balance between 
numerical accuracy and computation time. With this EOS, we are able to reliably model
stars with masses above the hydrogen burning minimum mass \citep{IrwinFEOSV}.

Surface boundary conditions are prescribed using {\sc phoenix ames-cond}
model atmospheres \citep{Hauschildt1999a}. Atmosphere structures are used
to define the initial gas pressure for our model envelope integration. 
Above $0.2\msun$, the gas pressure is determined an optical depth
where $T = \teff$. However, below $0.2\msun$, the regime where convection
is sufficiently non-adiabatic extends deeper into the
star \citep{CB97}. Thus, we specify our boundary conditions at the optical 
depth $\tau = 100$ in this mass regime. For the present work, we have extended
the initial metallicity grid of model atmosphere structures \citep{Dotter2007,Dotter2008}
by interpolating within the original set of structures. Care was taken to
ensure that the interpolation produced reliable results and that no discontinuities
in either $P_{\rm gas}$ or the starting temperature were introduced. This will be
discussed in a future publication. A set of atmosphere structures with finer metallicity
spacing allows for more accurate predictions of stellar properties at metallicities
that lie between the original grid spacings.

\subsection{Dynamos \& Radial Profiles}
\label{sec:rad_prof}

Implementation of a magnetic perturbation is described in detail by 
\citet{LS95} and \citet{FC12b}. We abstain from providing a mathematical 
description and refer the reader to those papers. However, it is beneficial 
to review multiple variations on our basic formulation that arose in 
\citet{FC13}. These variations take the form of 
different magnetic field strength radial profiles and what we have called 
different ``dynamos.'' The latter refers not to a detailed dynamo treatment, 
but a conceptual framework that concerns from where we assume the magnetic field 
sources its energy.

\subsubsection{``Rotational'' versus ``Turbulent'' Dynamo}

Our treatment of magneto-convection depends on how we assume the magnetic
field is generated. Assuming that rotation drives the dynamo, as in a standard
shell dynamo \citep{Parker1979}, leads to perturbations consistent with the idea that 
magnetic fields can stabilize a fluid against thermal convection \citep[e.g.,][]{
Thompson1951,Chandrasekhar1961,GT66}. This assumption forms the basis of
our magneto-convection formulation \citep{FC12b,FC13}. However, permissible 
magnetic field strengths can reach upward of 6\,kG at the model photosphere
with exact upper limits determined by equipartition with the thermal gas pressure.
This also results in interior magnetic field strengths that can grow nearly
without limit owing to large internal gas pressures. Models that rely on stabilizing
convection with a magnetic field are hereafter referred to as having ``rotational
dynamos.''

The form of the perturbation used for a star with an assumed rotational dynamo
is not necessarily valid in all stellar mass regimes. This is particularly true
in fully convective stars, where the interface region between the radiative core
and convective envelope is thought to disappear. To address this, and the problem
that magnetic fields in the rotational dynamo can grow without limit, we introduced
a ``turbulent dynamo'' mechanism \citep{FC13}. This formulation assumes
that the magnetic field strength at a given grid point within the model 
receives its energy from the kinetic energy of convecting material. Therefore,
the local Alfv\'{e}n velocity can not exceed the local convective velocity. It
is a simple approach developed to address zeroth-order effects within the 
already simplified convection framework of mixing length theory. 
For low-mass stars, this methods places an upper limit of around 3\,kG for 
surface magnetic field strengths, consistent with observed upper limits of average surface
magnetic field strengths \citep{Reiners2007,Shulyak2011}. Although rotation is 
still required for turbulent dynamo action \citep{Durney1993,Dobler2006,Chabrier2006,
Browning2008}, magnetic field strengths are more sensitive to properties of 
convection. 

We note, again, that the term ``dynamo'' is used loosely. Our formulations
of magneto-convection do not rigorously solve the equations of magnetohydrodynamics.
Instead, we seek to capture physically relevant effects on stellar structure 
in a phenomenological manner consistent with actual dynamo processes. Each
of the above descriptions rely equally on a prescribed magnetic field strength 
profile within the star.

\subsubsection{Dipole Radial Profile}

Models of the ``dipole radial field'' variety are the standard sort
introduced in \citet{FC12b}. This profile is characteristic of a magnetic 
field generated by a single current loop centered on the stellar tachocline. 
Given a surface magnetic field
strength, the radial profile of the magnetic field is determined by calculating
the peak magnetic field strength at the tachocline. Lacking a tachocline,
we define fully convective stars to have a peak magnetic field strength 
at 15\% of the stellar radius ($0.15R_\star$). This is loosely based on 
results from three-dimensional magnetohydrodynamic (MHD) models of 
fully convective stars \citep{Browning2008}. MHD models indicate that 
the magnetic field reaches a maximum around $\sim0.15 R_{\star}$ 
\citep{Browning2008}. Note that this maximum is not a sharply defined peak 
in the magnetic field strength profile since the profile is based largely
on equipartition of the magnetic field with convective flows.
Still, we adopt $0.15R_{\star}$ knowing 
this caveat, which we address in a moment. The rest of the interior
magnetic field is then calculated by assuming the magnetic
field strength falls off steeply towards the core and surface of the star.
Explicitly,
\renewcommand{\arraystretch}{1.5}
\begin{equation}
    B(R) = B_{\rm surf}\cdot\left\{
    \begin{array}{l l}
        R^3/R_{\rm tach}^6 & R < R_{\rm tach} \\
        R^{-3}             & R > R_{\rm tach}
    \end{array}
    \right.
\end{equation}
where $B_{\rm surf}$ is the prescribed surface magnetic field strength, 
$R_{\rm tach}$ is the radius of the tachocline normalized to the total stellar 
radius, and $R$ is the radius within the star normalized to the total stellar 
radius.

\subsubsection{Gaussian Radial Profile}

To increase the peak magnetic field strength for a given surface
magnetic field strength, we implemented a Gaussian profile 
\citep[Section 4.4.1 in][]{FC13}. The peak magnetic field strength is 
still defined at the tachocline in partially convective stars and at 
$R = 0.15 R_{\star}$ in fully convective stars. However, instead of a 
power-law decline of the interior magnetic field strength from the peak, 
the peak was set as the center of a Gaussian distribution. Thus,
\begin{equation}
    B(R) = B(R_{\rm tach}) \exp\left[ -\frac{1}{2}\left( \frac{R_{\rm tach} - R}{\sigma_{g}}\right)^2 \right],
\end{equation}
where $\sigma_g$ controls the width of the Gaussian and $B(R_{\rm tach})$
is defined with respect to the surface magnetic field strength. 
The width of the Gaussian depends on the depth of the convection zone. 
Deeper convection zones have a wider Gaussian profile compared to stars
with thin convective envelopes \citep{FC13}. Hereafter, we will refer to 
these as ``Gaussian radial profile'' models. Note that this prescription 
has no physical motivation beyond providing stronger interior magnetic 
fields.

\subsubsection{Constant $\Lambda$ Radial Profile}
\label{sec:clambda}
We define a third radial profile for this study: a ``constant $\Lambda$
radial profile.''  When a turbulent dynamo is invoked, the magnetic field
radial profiles described above can cause the Alfv\'{e}n velocity to exceed
the convective velocity within the model interior. This happens quite
easily in models of fully convective stars where the peak magnetic field
strength is defined deep within the model. As a result, perturbed convective
velocities become imaginary leading convergence problems when solving the 
equations of mixing length theory.

If we assume that kinetic energy in convective flows generates
the local magnetic field, an Alfv\'{e}n velocity exceeding the convective
velocity will lead to a decaying magnetic field strength. Eventually,
equipartition will be reached. We avoid iterating to a solution by using 
a profile that assumes a constant ratio of magnetic field to the
equipartition value, $B(r) = \Lambda B_{\rm eq}$, where $B_{\rm eq} = (4\pi\rho\uconv^2)^{1/2}$. 
This factor, $\Lambda$, was introduced in \citet{FC13} as a means of 
comparing reduced mixing length models \citep[i.e.,][]{Chabrier2007} to 
our turbulent dynamo models.

Perturbations to the equations of mixing length theory are now expressed
as functions of $\Lambda$. Removing energy from convection slows convective
flows such that
\begin{equation}
    \uconv = u_{\rm conv,\,0}\left(1 - \Lambda^2\right)^{1/2},
\end{equation}
where $u_{\rm conv,\,0}$ is the convective velocity prior to losing energy
to the magnetic field. We restrict $0 \le \Lambda \le 1$ to avoid imaginary
convective velocities. Reduction of convective velocity causes a significant
reduction in convective energy flux since $\mathcal{F} \propto \uconv^3$.
Steeping of the background temperature gradient, $\tgrad$, occurs as radiation 
attempts to transport additional energy. This increase, $\Delta\tgrad$, 
over the non-perturbed temperature gradient is 
\begin{equation}
    \Delta\tgrad = \frac{(\Lambda u_{\rm conv\, 0})^2}{C},
\end{equation}
where $C = g\amlt^2H_P\delta/8$ is the characteristic squared velocity of
an unimpeded convecting bubble over a pressure scale height. This is a sort 
of terminal convective velocity, where $g$ is the local acceleration due
to gravity, $\amlt$ is the convective mixing length parameter, $H_P$ is
the pressure scale height, and $\delta = (\partial\ln\rho/\partial\ln T)_{P,\,\chi}$ 
is the coefficient of thermal expansion. Resulting effects on convection 
likely represent an upper limit on the effects of such a dynamo mechanism 
in inhibiting thermal convection. Note that when using this radial profile
and dynamo mechanism, modifications to the Schwarzschild criterion
\citep[i.e., the formalism outlined in][]{FC12b} are neglected.

\section{Analysis of Individual DEB Systems}
\label{sec:ind_deb}
We have chosen to study the DEBs Kepler-16 and CM Dra in detail. LSPM
J1112+7626 was not modeled in detail because it lacks a proper metallicity
estimate required for a careful comparison with models. Kepler-38 was
also not selected as initial comparisons suggest that there may be some
issues with modeling the primary. Finally, KOI-126 has been modeled in
detail previously and does not appear to require magnetic fields 
\citep{Feiden2011,SD12}. As we will show below, magnetic models of Kepler-16~B
and CM Dra produce similar results that can be generalized to all 
fully convective main sequence stars in DEBs.

\subsection{Kepler-16}
\label{sec:kep16}

\begin{deluxetable}{l c c}
    \tablecolumns{3}
    \tablecaption{Fundamental properties of Kepler-16.}
    \tablewidth{0.85\linewidth}    
    \tablehead{
    Property  &  \colhead{Kepler-16 A}  &  \colhead{Kepler-16 B}
    }
    \startdata
        D11 Mass ($\msun$)        & $0.6897 \pm 0.0034$  & $0.20255 \pm 0.00066$ \\
        B12 Mass ($\msun$)        & $0.654  \pm 0.017$   &  $0.1959 \pm 0.0031$  \\
        Radius ($\rsun$)          & $0.6489 \pm 0.0013$  &  $0.2262 \pm 0.0005$  \\
        $\teff$ (K)               &   $4337 \pm\ 80$     &        $\cdots$       \\ 
        $\left[{\rm Fe/H}\right]$ & \multicolumn{2}{c}{$-0.04\pm0.08$}       \\
        Age (Gyr)                 &    \multicolumn{2}{c}{$3\pm1$}            
    \enddata
    \label{tab:kep_16_prop}
\end{deluxetable}

\begin{figure*}
    \plottwo{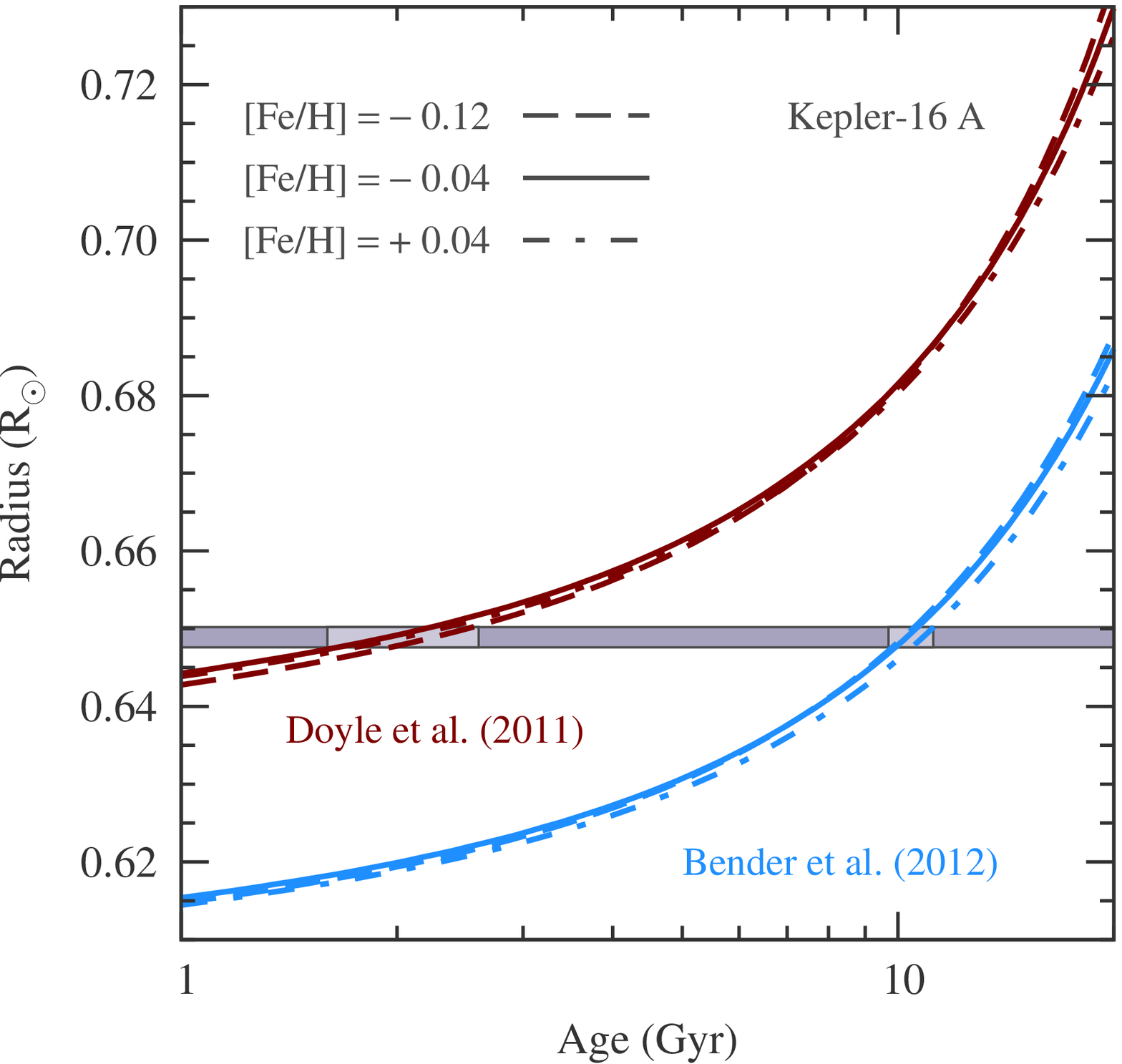}{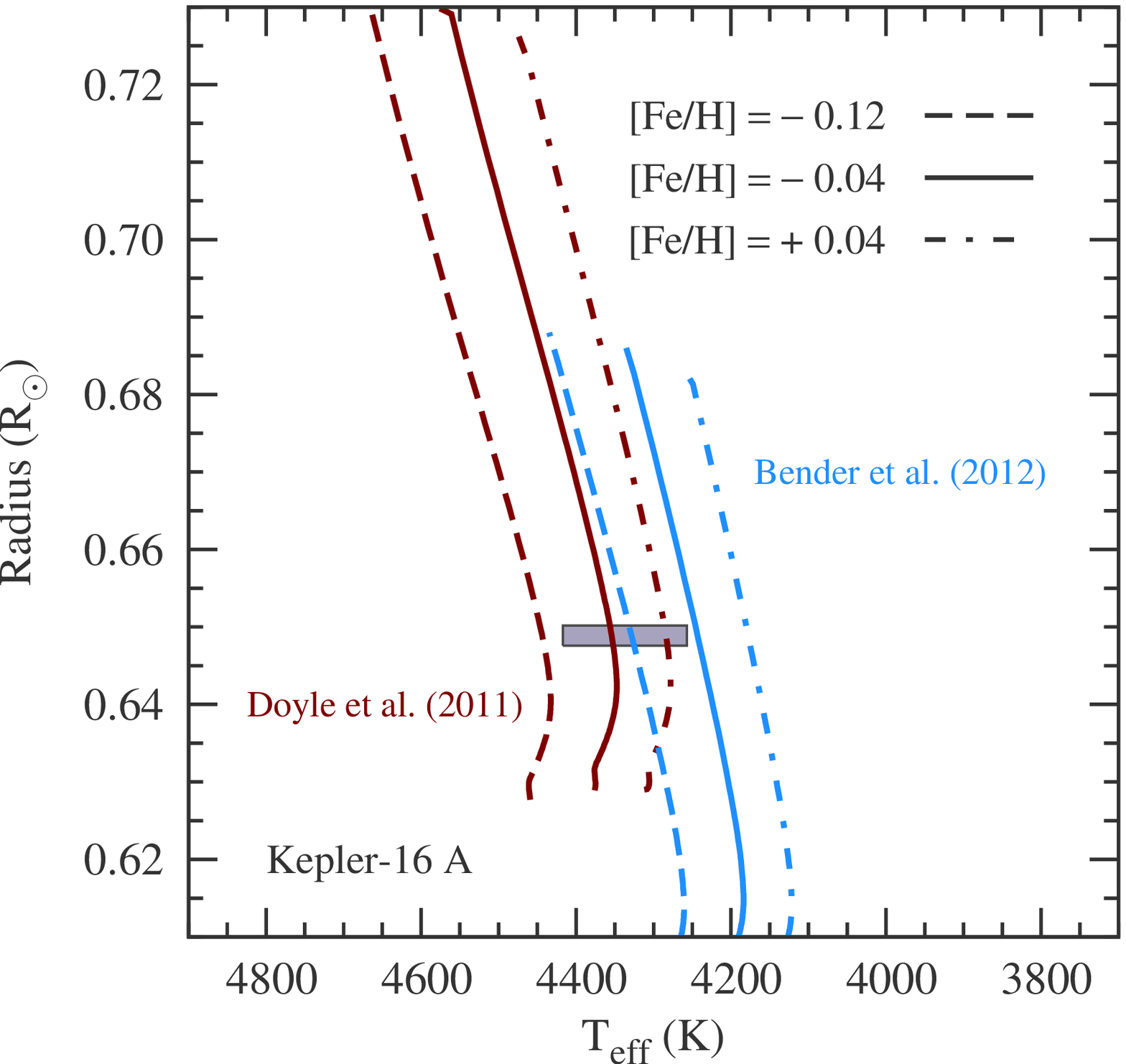}
    \caption{Standard Dartmouth models computed at the exact masses measured
     by \citet{Doyle2011} (maroon) and \citet{Bender2012} (light-blue) 
     for Kepler-16 A. Mass tracks for the adopted metallicity of \citet{Winn2011}
     and the two limits of the associated $1\sigma$ uncertainty are 
     given by solid, dash, and dash-dotted lines, respectively. (a) 
     Age-radius diagram with the observed radius indicated by the purple 
     horizontal swath. (b) $\teff$-radius plane where the purple box indicates 
     observational constraints for Kepler-16 A.
    }
    \label{fig:kep16_a_nmag}
\end{figure*}

Report of the first circumbinary exoplanet was announced by 
\citet{Doyle2011} with their study of Kepler-16b. While the planet is 
interesting in its own right, what made the finding even more interesting 
was that the two host stars formed a long-period low-mass DEB. This enabled a precise 
characterization of the stars and circumstellar environment. Kepler-16 
contains a K-dwarf primary with a fully-convective M-dwarf secondary
in a 41 day orbit. Properties of the two stars are listed in Table 
\ref{tab:kep_16_prop}.

In a follow-up investigation, \citet{Winn2011} estimated the composition
and age of the system. An age of $3\pm1$\,Gyr was estimated using
gyrochronology and an age-activity relation based on Ca {\sc ii} emission.
Spectroscopic analysis of the primary revealed the system has a solar-like
metallicity of [Fe/H] $= -0.04\pm0.08$\,dex. They also compared properties
of the Kepler-16 stars to predictions of stellar evolution models. \citet{BCAH98}
model predictions agreed with properties of the primary at the given age. However, the 
radius of the fully convective secondary star was larger than model predictions
by $\sim3\%$. Independent confirmation of the stellar properties and model
disagreements was provided by \citet{FC12}, who found a best fit age of 
1 Gyr with [Fe/H] $= -0.1$ using Dartmouth models.

Radial velocity confirmation of the component masses---within $2\sigma$---was 
later obtained by \citet{Bender2012}. They found masses that were 5\% 
lower than the original masses \citep{Doyle2011}. Of particular note, is  
that mass ratio is different between the two studies. \citet{Bender2012} 
attempted to pin-point the origin of this discrepancy, but were unable to 
do so with complete confidence. 

Despite the disagreement, the spectroscopic masses largely confirm
that masses derived using a photometric-dynamical model are reliable 
\citep{Carter2011,Doyle2011}. However, a slight mass difference significantly 
alters comparisons with stellar evolution models. That is, if we assume 
that the derived masses do not heavily influence the radius and effective 
temperature predictions. We therefore opt to treat the two different mass 
estimates independently to assess how these differences affect our modeling
efforts. Hereafter, masses originally quoted by \citet{Doyle2011} 
will be referred to as \citetalias{Doyle2011} masses, whereas the revised 
values of \citet{Bender2012} will be referred to as \citetalias{Bender2012} 
masses. Table \ref{tab:kep_16_prop} lists masses measured by each group.

\subsubsection{Standard Models}

We first focus our attention on Kepler-16 A, as the analysis will
directly influence our analysis of the secondary. Plotted in
Figure~\ref{fig:kep16_a_nmag} are non-magnetic Dartmouth models computed 
at the measured masses of the primary provided by \citetalias{Doyle2011} 
and \citetalias{Bender2012}. Three separate 
tracks are illustrated for each mass estimate, corresponding to [Fe/H] 
$= -0.12, -0.04$, and $+0.04$ \citep{Winn2011}. In this figure, the horizontal 
shaded region highlights the observed radius with associated $1\sigma$ 
uncertainties. Ages are determined by noting where the mass tracks are
located within the bounds of the empirical radius constraints. Similarly,
we confirm that when the mass track has the required radius it also has 
an appropriate effective temperature in Figure \ref{fig:kep16_a_nmag}(b).

Given a primary mass from \citetalias{Doyle2011}, we find that standard  
stellar evolution models match the stellar radius and temperature at an 
age of $2.1\pm0.5$\,Gyr. When the model radius equals the precise empirical 
radius ($0.6489\rsun$), the associated model effective temperature is 4354 K,
which is within 17\,K of the spectroscopic effective temperature \citep{Winn2011}.
We note that this age is also consistent with the estimated age from 
\citet{Winn2011}.  

Agreement between models and observations does not guarantee the validity 
of the \citetalias{Doyle2011} masses over the \citetalias{Bender2012} 
masses. In fact, this agreement is rather expected. The effective temperature 
and metallicity for the primary were determined using Spectroscopy Made 
Easy \citep[hereafter SME]{sme}, which relies on theoretical stellar 
atmospheres. Our models also rely on theoretical atmospheres, although
{\sc phoenix} model atmospheres, used by our models, adopt a different line 
list database \citep[see][and references therein]{Hauschildt1999a} 
than the theoretical atmospheres used by SME \citep[VALD: Vienna Atomic 
Line Database;][]{Piskunov1995}. In a sense, the fact that we 
find such good agreement with the effective temperature for a given $\log g$ and metallicity 
(i.e., those of Kepler-16 A) may be a better test of the agreement between
different stellar atmosphere models rather than a test of the interior 
evolution models. What is encouraging, is that we derive appropriate
stellar properties at an age consistent with gyrochronology and age-activity
relations \citet{Winn2011}. Age consistency is not ensured by agreement 
between stellar atmosphere structures.

Our previous discussion may be erroneous if we adopt an incorrect 
mass. \citetalias{Bender2012} suggest this is the case. The effect of 
adopting the lower B12 masses is displayed in Figure~\ref{fig:kep16_a_nmag}. 
Assuming the radius measurement remains constant, we derive an age of 
$10.5\pm0.8$\,Gyr 
for the primary star. The effective temperature associated with the model 
also appears too cool compared to observations at the measured metallicity 
($-0.04$\,dex). Relief is found by lowering the metallicity by $0.1$\,dex, 
which increases the temperature by $30$\,K. This is enough to bring the 
model temperature to within $1\sigma$ of the spectroscopic value \citet{Winn2011}.

There is a caveat: the spectroscopic analysis by \citet{Winn2011} relied 
on fixing the stellar $\log g$ as input into SME. Thus, the temperature 
and metallicity are intimately tied to the adopted $\log g$. Reducing the 
mass of the primary by 5\%, as is done by \citetalias{Bender2012}, but 
leaving the radius fixed to the \citetalias{Doyle2011} value leads to a 
decrease in $\log g$ of $0.02$ dex. Such a change in the fixed value of 
$\log g$ may decrease the derived effective temperature and bring the
model and empirical temperatures into agreement. All things considered,
we believe it is safe to assume that a shift in mass does not introduce
any significant effective temperature disagreements at a given radius.
This is predicated on the fact that the temperature is safely above $\sim4000$\,K.
Below this temperature, theoretical atmosphere predictions start to degrade
with the appearance of molecular bands. More simply stated, we have no 
reason to doubt the Dartmouth model predicted temperature for Kepler-16 A, 
regardless of the adopted mass.

A model age of 11\,Gyr for the \citetalias{Bender2012} primary mass appears 
old given the multiple age estimates provided by \citet{Winn2011}. Is it 
possible that the system is actually 11\,Gyr old, but appears from rotation 
and activity to be considerably younger? Consider that a $35.1\pm1.0$\,day 
rotation period of the primary, as measured by \citet{Winn2011}, is nearly
equal to the pseudo-synchronization rotation period, predicted to be 35.6\,days 
\citep{Hut1981,Winn2011}. One might think that tidal effects are unimportant 
in a binary with a 41 day orbital period, however, subsequent tidal 
interactions briefly endured when the components are near periastron 
can drive the components towards pseudo-synchronous rotation. 
Pseudo-synchronous rotation will keep the stars rotating at a faster rate 
than if they were completely isolated from one another. The time scale for 
this to occur is approximately 3\,Gyr \citep{Winn2011}, meaning that the 
rotation period is not necessarily indicative of the system's age. The 
primary will have approximately the same rotation period at 11\,Gyr as it 
will at 3\,Gyr. Furthermore, the timescale for orbital circularization is 
safely estimated to be between $\sim10^{4}$ -- $10^{5}$Gyr \citep{Winn2011}. Tidal 
evolution calculations are subject to large uncertainties and should 
approached as an order of magnitude estimate. But, we are unable to 
immediately rule out the possibility that Kepler-16 has an age of 11\,Gyr.

\begin{figure}[t]
    \plotone{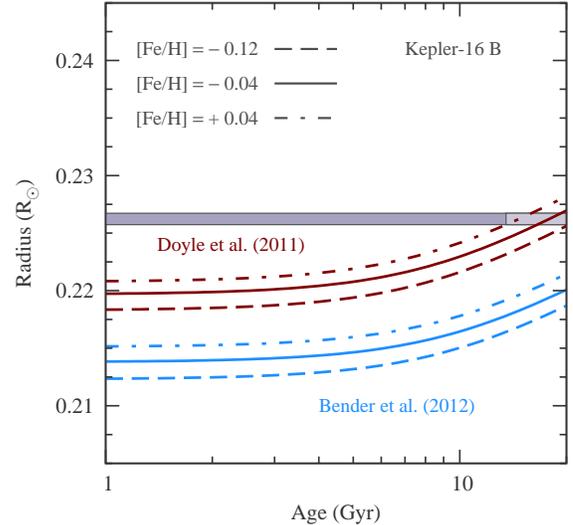}
    \caption{Identical to Figure \ref{fig:kep16_a_nmag}(a), except the 
        mass tracks are computed at the masses measured for Kepler-16 B.}
    \label{fig:kep16_b_nmag}
\end{figure}

We have so far neglected any remark on the agreement between standard 
models and Kepler-16 B. This comparison is carried out in Figure 
\ref{fig:kep16_b_nmag}. No effective temperature estimates 
have been published, explaining our neglect of the $\teff$-radius plane. 
As with Figure~\ref{fig:kep16_a_nmag}, mass tracks are shown
for multiple metallicities. Standard model mass tracks for Kepler-16 
B are unable to correctly predict the observed radius at an age consistent
with estimates from the primary. 
The disagreement is independent of the adopted metallicity, 
which introduces $\sim0.5\%$ variations in the stellar 
radius at a given age.

No evidence is available to support the idea that Kepler-16 B is
magnetically active. Still, we look to magnetic fields to reconcile the 
model predictions with the observations. All possible scenarios relating 
to the various stellar mass estimates are considered. Explicitly, we 
compute models for both the \citetalias{Doyle2011} and \citetalias{Bender2012} 
masses and then attempt to fit 
the observations using the magnetic Dartmouth stellar evolution models.

\begin{figure}[t]
    \plotone{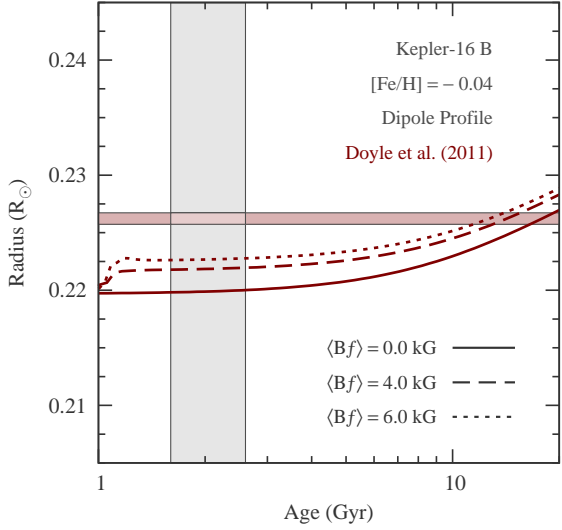}
    \caption{Standard (solid line) and magnetic (broken lines) Dartmouth 
     models of Kepler-16 B with a \citetalias{Doyle2011} mass.  
     Models were computed with [Fe/H] $= -0.04$ and a solar calibrated 
     $\amlt$. Magnetic models were calculated using a dipole radial 
     profile with a $4.0$\,kG (dashed line) and a $6.0$\,kG (dotted
     line) surface magnetic field strength. Observed radius constraints 
     are shown as a shaded horizontal region and an age constraint is given
     by the vertical shaded region.}
    \label{fig:kep16b_dip}
\end{figure}

\subsubsection{Magnetic Models: D11 Masses}

The \citetalias{Doyle2011} primary mass implies that Kepler-16 is approximately 
2\,Gyr old, as shown in Figure~\ref{fig:kep16_a_nmag}(a). Due to the 
consistency between the model derived age and the empirically inferred age, 
we see no reason to introduce a magnetic perturbation into models of 
Kepler-16 A. \citet{Winn2011} observe only moderately weak chromospheric 
activity coming from the primary, further supporting our decision. Thus,
we seek to reconcile models of Kepler-16 B with \citetalias{Doyle2011} 
masses at 2\,Gyr.

Magnetic models of Kepler-16 B were computed for a range of surface
magnetic field strengths. A dipole radial profile was used and the perturbation
was applied over a single time step at an age of 1\,Gyr. Mass tracks with 
a $4.0$\,kG and $6.0$\,kG surface magnetic field strength are shown in Figure 
\ref{fig:kep16b_dip} along side a standard mass track for comparison. Note, 
that even though the perturbation is applied at an age close to the age we
are trying to fit, the models adjust to the perturbation rapidly. We are 
unable to produce a radius inflation larger 
than 1\%, even with a strong surface magnetic field strength of $6.0$\,kG. 
The peak field strength in the $6.0$\,kG model (located at $R = 0.15\,R_{\star}$) 
is approximately 1.8\,MG ($\nu \approx P_{\rm mag}/P_{\rm gas} = 10^{-6}$). Discussion 
about how real such a magnetic field
might be is deferred until Section \ref{sec:mag_conv}. For the moment, we
are interested in knowing what magnetic field strength is required to
reconcile models with observations.

\begin{figure}[t]
    \plotone{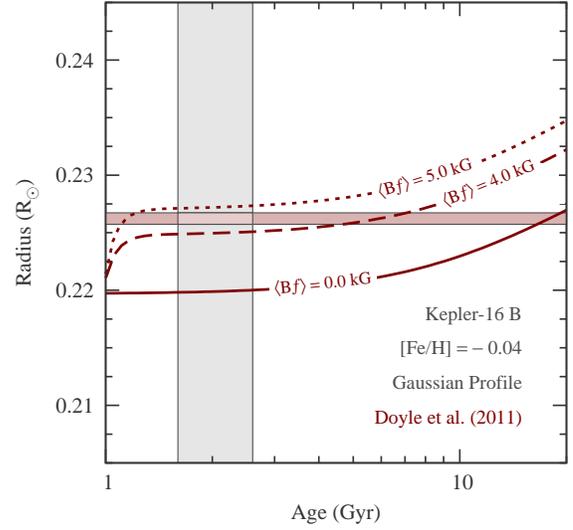}
    \caption{Same as Figure \ref{fig:kep16b_dip}, except the magnetic models
     were computed with a Guassian radial profile.}
    \label{fig:kep16b_d11}
\end{figure}

We next constructed magnetic models using a Gaussian radial profile, which
are shown in Figure  \ref{fig:kep16b_d11}. The magnetic perturbation was 
again introduced as a single perturbation at 1\,Gyr. Two surface 
magnetic field strengths were used, $4.0$\,kG and $5.0$\,kG. The model 
with a $5.0$ kG surface magnetic field strength causes the model to become
over inflated compared to the observed radius at 2\,Gyr. From Figure 
\ref{fig:kep16b_d11}, we see that a magnetic field intermediate between
4\,kG and 5\,kG is required to produce agreement between the models and 
observations.
In contrast to results for partially convective stars \citep{FC13}, dipole 
and Gaussian radial profiles produce different results for a given surface 
magnetic field strength in fully convective stars. This is caused by a 
difference in peak magnetic field strengths (dipole: 1.8\,MG, $\nu = 10^{-6}$; 
Gaussian: 30\,MG, $\nu = 10^{-4}$). We will return to this issue in Section 
\ref{sec:fc_profs}. 

\begin{figure}
    \plotone{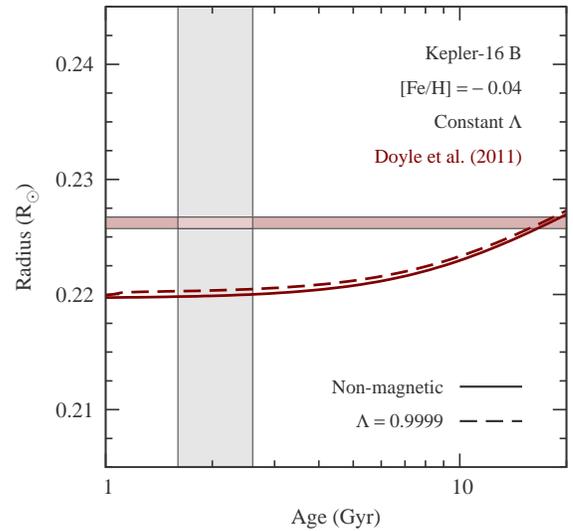}
    \caption{Same as Figure \ref{fig:kep16b_dip}, but the magnetic model 
     was calculated using a constant $\Lambda = 0.9999$ profile.}
    \label{fig:kep_16_beq}
\end{figure}

Finally, Figure \ref{fig:kep_16_beq} shows the influence of a constant 
$\Lambda$ profile on a model of Kepler-16 B. Recall, this radial profile
invokes a turbulent dynamo mechanism. We started with a rather 
high value of $\Lambda = 0.9999$ to gauge the model's reaction to this 
formulation. Impact on the radius evolution of a $M = 0.203\msun$ star 
is effectively negligible. Further increasing $\Lambda$ has no effect 
on the resulting radius evolution.

\subsubsection{Magnetic Models: B12 Masses}

Adopting \citetalias{Bender2012} masses mainly alters the age derived from stellar 
models. Instead of 2\,Gyr, we infer an age of 11\,Gyr from models of 
Kepler-16 A, as was shown in Figure \ref{fig:kep16_a_nmag}(a). The relative 
radius discrepancy noted between models and Kepler-16 B is increased by 
approximately 2\% over the \citetalias{Doyle2011} case. 

\begin{figure}
    \plotone{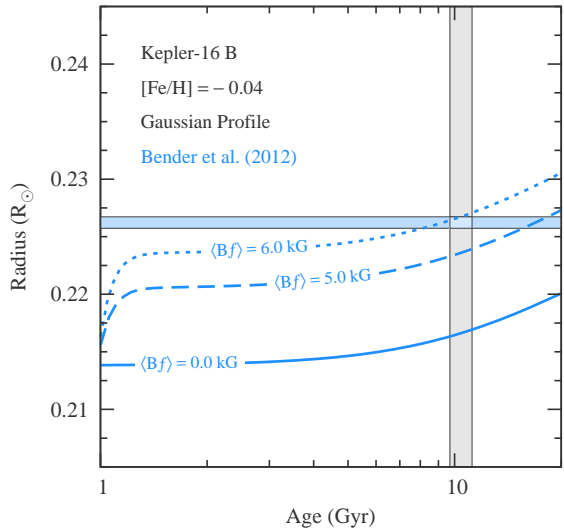}
    \caption{Similar to Figure \ref{fig:kep16b_d11}, but with
     \citetalias{Bender2012} masses and a Gaussian radial profile. Surface
     magnetic field strengths used were $5.0$\,kG (dotted line) and $6.0$\,kG
     (dash-dotted line).}
    \label{fig:kep16b_b12}
\end{figure}

Magnetic models were computed for Kepler-16 B with the \citetalias{Bender2012} 
mass estimate 
using a Gaussian radial profile introduced at an age of 1\,Gyr. These models
are shown in Figure \ref{fig:kep16b_b12}. We did not generate models with
a dipole radial profile or with a constant $\Lambda$ profile given the 
lack of radius inflation observed for these magnetic field profiles for 
the \citetalias{Doyle2011} masses. 
We find that a surface magnetic field strength slightly weaker than $6.0$\,kG 
is required to fit the observations. This
translates to a nearly $40$\,MG ($\nu = 10^{-4}$) peak magnetic field strength. 
A stronger 
field strength was required when using the \citetalias{Bender2012} mass 
instead of the \citetalias{Doyle2011} mass 
because of the 1\% increase in the radius discrepancy mentioned above.

\subsubsection{Summary}

Kepler-16, although it has components with fundamental properties measured 
with better than 3\% precision, must be approached cautiously when comparing 
to stellar models. Mass differences quoted in the literature obscure how 
well models perform against the observations. Though the masses are determined
with high precision by each group, the 3\% -- 5\% uncertainty introduced by their 
disagreement overwhelms the measurement precision. This produces an age 
difference of 9\,Gyr for the primary star, calling into question inferences 
drawn about this system from stellar models. It is also unclear how revising
the masses would affect estimates of the component radii, the system's metallicity,
and the primary star's effective temperature. Furthermore, since the primary
may be rotating pseudo-synchronously, it is not possible to rule out either
of the age estimates. 

Until the mass differences are resolved, {\it care must be taken when comparing
Kepler-16 to stellar models}. However, the disparity between the observed
radius of Kepler-16 B and model predictions is apparent regardless of the
adopted mass estimate. Changing the mass estimate simply changes the level
of inferred radius inflation. This result appears robust and can be used to test the 
magnetic hypothesis for low-mass star radius inflation. Our magnetic models 
require magnetic field strengths of similar magnitudes. Surface magnetic
field strengths are on the order of 4\,kG -- 6\,kG with interior field 
strengths of a few tens of MG. 

\subsection{CM Draconis}
\label{sec:cmdra}

CM Dra (GJ 630.1 AC) contains two fully convective low-mass stars and
is arguably one of the most important systems for benchmarking stellar 
evolution models. Shortly after CM Dra 
was discovered by Luyten, \citet{Eggen1967} uncovered that the star
was actually a DEB. It was not clear from observations whether the secondary
was a dark, very low-mass companion such that no secondary eclipse occurred
or whether the two components were of nearly equal mass. Evidence was tentatively
provided in favor CM Dra being a single dMe star with a dark companion
\citep{Martins1975}, although more observations were encouraged as the 
author found a possible hint of a secondary eclipse. Any speculation that 
the secondary companion to the dMe star of CM Dra was
a dark, lower-mass object was laid to rest by \citet{Lacy1977} who obtained 
radial velocity measurements to provide the first determination of stellar parameters 
for both stars.

\begin{deluxetable}{l c c}[b]
    \tablecolumns{3}
    \tablecaption{Fundamental properties of CM Draconis.}
    \tablewidth{0.85\linewidth}
    \tablehead{
        Property  &  \colhead{CM Dra A}  &  \colhead{CM Dra B}
    }
    \startdata
        Mass ($\msun$)            & $0.23102\pm0.00089$  & $0.21409\pm0.00083$ \\
        Radius ($\rsun$)          & $0.2534\pm0.0019$    & $0.2398\pm0.0018$   \\
        $\teff$ (K)               &    $3130\pm\ 70$     &   $3120\pm\ 70$     \\ 
        $\left[{\rm Fe/H}\right]$ & \multicolumn{2}{c}{$-0.30\pm0.12$}     \\
        Age (Gyr)                 &    \multicolumn{2}{c}{$4.1 \pm 0.8$}       
    \enddata
    \label{tab:cm_dra_prop}
\end{deluxetable}

\begin{figure}[b]
    \plotone{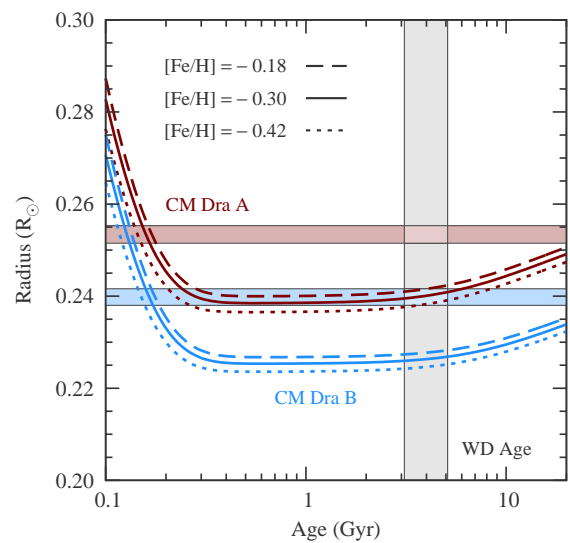}
    \caption{
     Standard Dartmouth models of CM Dra A (maroon) and B (light-blue). 
     Models were computed with [Fe/H] $= -0.18$ (dashed line), $-0.30$
     (solid line), and $-0.42$ (dotted line) and a solar calibrated 
     $\amlt$. Observed radius constraints are shown as shaded horizontal 
     regions and an age constraint is given by a vertical shaded region.}
    \label{fig:cmdra_nmag}
\end{figure}

\begin{figure}[t]
    \plotone{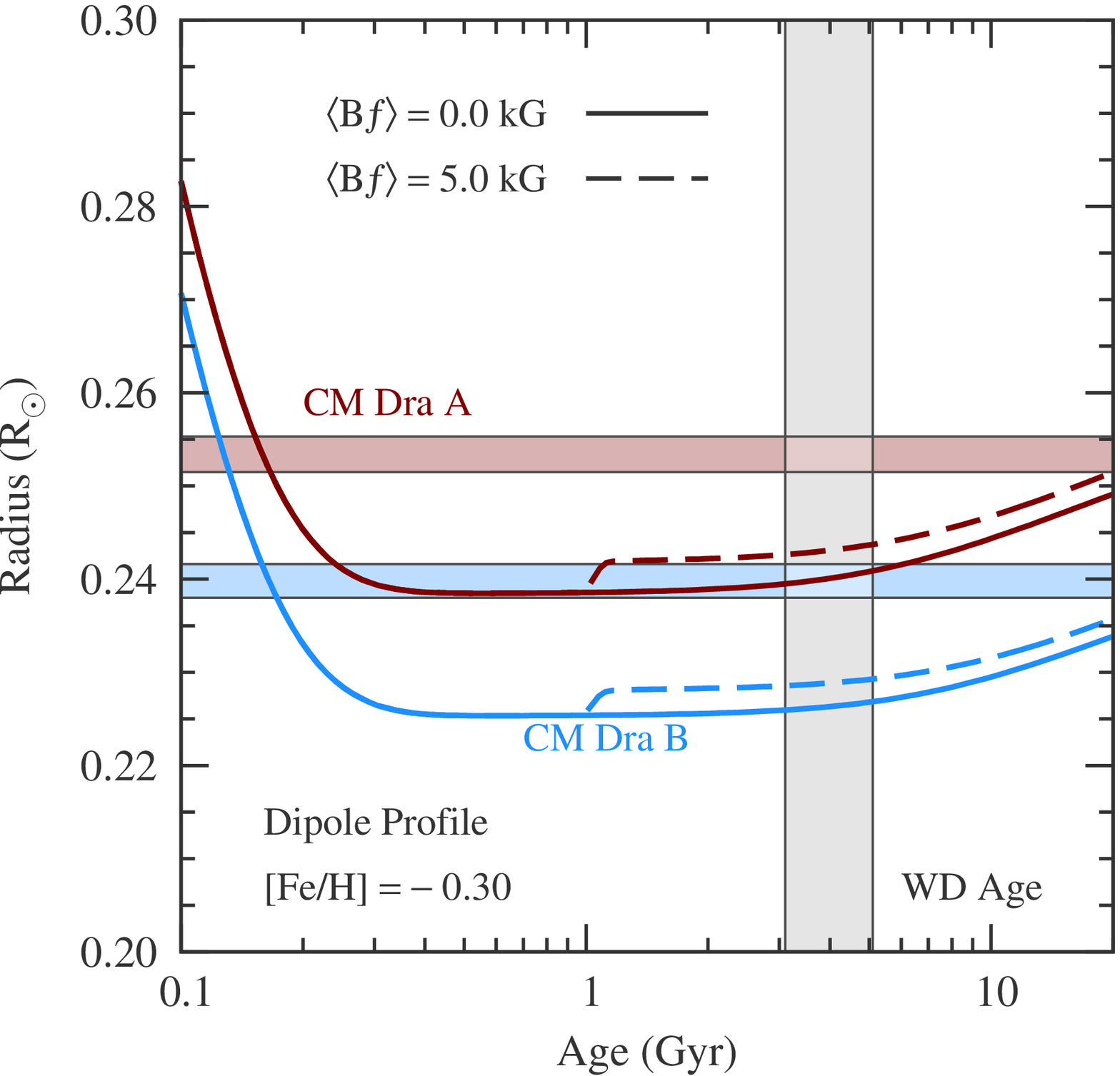}\\
    \vspace{\baselineskip}
    \plotone{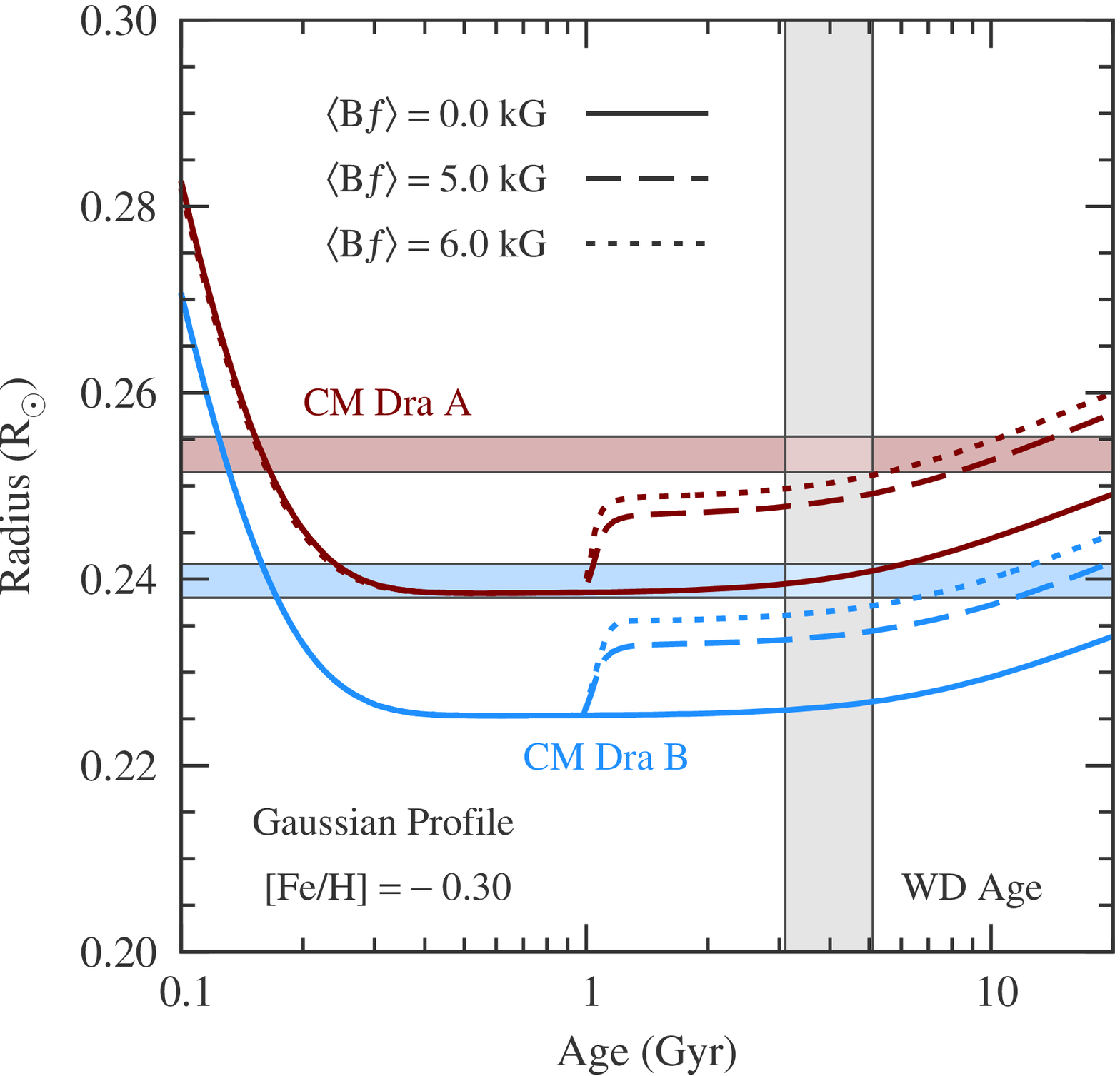}
    \caption{
     Standard (solid line) and magnetic (dashed line) Dartmouth models of 
     CM Dra A (maroon) and B (light-blue). Models were computed with 
     [Fe/H] $=-0.18$ and a solar calibrated $\amlt$. (top) Magnetic models
     with a dipole radial profile and a $5.0$\,kG surface magnetic field
     strength. (bottom) Magnetic models with a Gaussian radial profile and
     a $6.0$\,kG surface magnetic field strength. Observed radius constraints 
     are shown as shaded horizontal regions (color-matched to the mass 
     tracks) and an age constraint is given by a vertical shaded region.}
    \label{fig:cmdra_mag_d}
\end{figure}

Following Lacy's determination of the stellar properties, subsequent studies 
refined and improved the masses and radii of the CM Dra stars, pushing the 
measurement precision below 2\% \citep{Metcalfe1996,Morales2009a}. Currently 
accepted values \citep{Morales2009a,Torres2010} are listed in Table~\ref{tab:cm_dra_prop}.
Additional information about the CM Dra stars has been revealed in recent 
years. \citet{Morales2009a} provided an analysis of a nearby white 
dwarf (\object{WD 1633+572}) common proper motion companion and estimated 
an age of 4.1 $\pm$ 0.8 Gyr. This age was based upon the cooling time of 
the white dwarf and its estimated progenitor lifetime, which depends on 
the initial to final  mass relation for white dwarfs.  In light of recent 
advances in our understanding of white dwarf cooling \citep[e.g.,][]{Salaris2010} 
and the initial-to-final mass relation \citep[e.g.,][]{Kalirai2009} we 
intend to re-examine the question of the age of CM Dra in a future paper.
For the present work, we adopt the aforementioned age.

Deriving a metallicity for CM Dra has proven more difficult than estimating
its age. Near-infrared (NIR) spectroscopic studies that fit theoretical 
model atmospheres to atomic and molecular features have consistently favored 
a metal-poor abundance
\citep[${\rm [M/H]} \approx -0.6$;][]{Viti1997,Viti2002,Kuznetsov2012}.
Optical spectroscopy of molecular features \citep[CaH \& TiO;][]{Gizis1997} 
and NIR photometric colors \citep{Leggett1998}, on the other hand,
suggests that the system might have a near-solar metallicity. More recent 
techniques relying on empirically-calibrated narrow-band NIR (H \& K band)
spectral features have converged on a value of [M/H] = $-0.3\pm0.1$
\citep{RojasAyala2012,Terrien2012}. Further support for the latter estimate
is provided by the photometric color-magnitude-metallicity relation of 
\citet{Johnson2009}, which predicts [Fe/H] $\approx -0.4$. For this study,
we adopt [Fe/H] = $-0.30\pm0.12$ presented by \citet{Terrien2012} who controlled
for uncertainties introduced by orbital phase variations.

\begin{figure}[t]
    \plotone{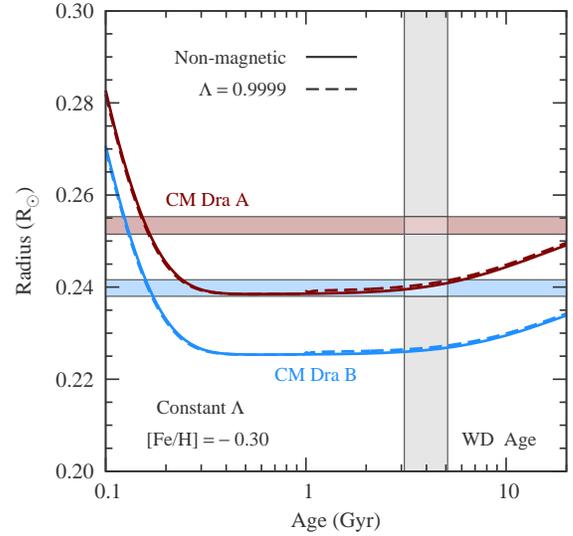}
    \caption{Same as Figures \ref{fig:cmdra_mag_d}(a) and (b), but the 
     magnetic model was 
     calculated with a constant $\Lambda=0.9999$ turbulent dynamo 
     formulation.}
    \label{fig:cmdra_mag_2}
\end{figure}

It is well documented that the stars of CM Dra are inflated compared to
standard stellar models \citep{Ribas2006,Morales2009a,Torres2010,FC12,SD12,Terrien2012}.
This fact has become steadily more apparent since an initial comparison
was performed by \citet{CB1995}, which found little disagreement. A precise 
estimate of the level of disagreement depends on the adopted metallicity 
\citep[see, e.g.,][]{FC12,Terrien2012}, but the problem is robust. Figure 
\ref{fig:cmdra_nmag} demonstrates the level of disagreement compared to 
standard Dartmouth mass tracks.It also illustrates how metallicity influences
the stellar models. Given that there is no evidence for polar spots on CM 
Dra (see Section \ref{sec:spots}), we have elected to use the radii established 
by \citet{Torres2010}. The level of disagreement observed in Figure \ref{fig:cmdra_nmag} 
is between 5\% -- 7\% for each star, with CM Dra A have a consistently smaller
deviation than CM Dra B by about 0.5\%. 
\citet{Terrien2012}, as a consequence of their metallicity estimate, have 
essentially doubled the radius disagreement from 3\% -- 4\%, noted in previous
studies \citep[e.g.,][]{FC12}, to 5\% -- 7\%.

CM Dra is magnetically active. Balmer emission
and light curve modulation due to spots have been recognized since very
early investigations \citep{Zwicky1966,Martins1975,Lacy1977}. Frequent
optical flaring has also been continually noted \citep[e.g.,][]{Eggen1967,
Lacy1977}. Further details on the flare characteristics of CM Dra may be
found in the work by \citet{MM11}. The system is also a strong source
of X-ray emission based on an analysis of data in the ROSAT All-Sky Survey 
Bright Source Catalogue \citep{Voges1999,Lopezm2007,FC12}.

High levels of magnetic activity and a short orbital period ($1.27$\,d) 
have been used to justify the need for magnetic perturbations in 
stellar evolution models of CM Dra. Such studies were carried out by 
\citet{Chabrier2007}, \citet{Morales2010}, and \citet{MM11} using various
methods (see Section \ref{sec:comp_study} of this work). In each case, 
magnetic models were found to provide satisfactory agreement with the 
observations. CM Dra therefore provides a pivotal test of our models 
and of the magnetic field hypothesis. 

Magnetic model mass tracks are displayed in Figures \ref{fig:cmdra_mag_d}
and \ref{fig:cmdra_mag_2}. We computed magnetic models using all three 
radial profiles discussed in Section \ref{sec:rad_prof} and using both 
dynamo formulations. Perturbations 
were introduced over a single time step at an age of 1\,Gyr. As with Kepler-16~B,
the magnetic models adjust to the perturbation well before the age where we 
attempt to perform the fit of models to observations. All magnetic models have 
[Fe/H] $= -0.30$ and a solar-calibrated $\amlt$. Since model radius predictions 
are only affected at the 1\% level due to metallicity variations, inflation 
required of magnetic fields is the dominating factor when attempting to correct
radius deviations of 6\%.

Results for CM Dra are similar to those for Kepler-16 B. Models with a dipole
radial profile are unable to inflate model at the level needed, despite strong
surface magnetic field strengths being applied (5.0\,kG -- 6.0\,kG). This is 
evidenced in Figure \ref{fig:cmdra_mag_d}(a), where magnetic-field-induced
inflation occurs at the 1\% -- 2\% level. Gaussian radial profile models were able to 
largely reconcile models with observations. Figure \ref{fig:cmdra_mag_d}(b) shows
that surface magnetic field strengths of $\sim6.0$\,kG are required to provide
the necessary radius inflation. Note that in \ref{fig:cmdra_mag_d}(b) the 6.0\,kG
mass track for CM Dra A actually has a 5.7\,kG magnetic field for reasons related
to model convergence. One can also see in Figure \ref{fig:cmdra_mag_d}(b) that 
CM Dra B requires a slightly stronger magnetic field strength. As with Kepler-16
B, the peak interior magnetic field strengths are 1.8\,MG ($\nu = 10^{-6}$) and 
40\,MG ($\nu = 10^{-4}$) for the dipole and Gaussian profiles, respectively.
Constant $\Lambda$ models are shown in Figure \ref{fig:cmdra_mag_2}. We only 
plot a $\Lambda = 0.9999$ for each mass. Model radius inflation induced by 
these models is negligible, as with the case for Kepler-16 B.

\section{Discussion}
\label{sec:disc}

\subsection{Magnetic Field Radial Profiles}
\label{sec:fc_profs}
The different results produced by the three magnetic field profiles introduced
in Section \ref{sec:rad_prof} can be understood in terms of convective
efficiency \citep{Spruit1986,FC13}. Convection near the surface of partially
convective stars displays a higher level of super-adiabaticity than does 
convection in the outer layers of fully convective stars. In general, this 
suggests that convection is less efficient in the outer layers of partially
convective stars. The structure of partially convective stars is therefore
more sensitive to changes in convective properties. As a result, structural 
changes induced by modification to convective properties at the stellar surface
induce the necessary radius inflation before the interior magnetic field strength
becomes large enough to inhibit convection near the base of the convection zone.
Dipole and Gaussian profiles then produce similar results for partially convective 
stars.

\begin{figure}[t]
    \plotone{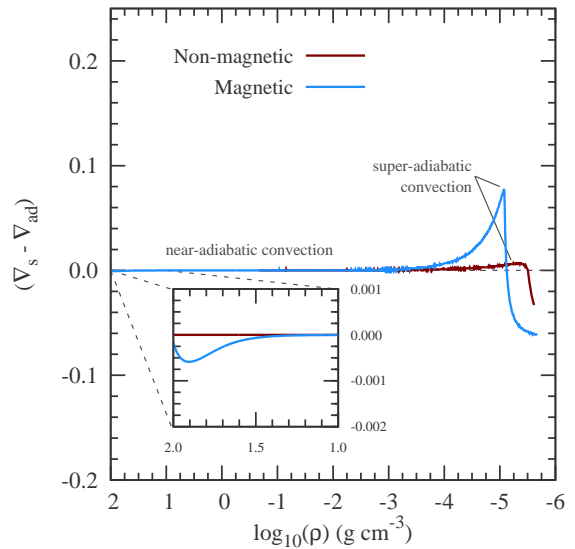}
    \caption[($\tgrad-\dela$) as a function of the logarithmic density.]
     {The difference between the real temperature gradient,
     $\tgrad$, and the adiabatic temperature gradient, $\dela$, as a
     function of the logarithmic plasma density for a $M=0.231\msun$ star. 
     We show this for two
     models: a non-magnetic model (maroon, solid line) and a magnetic 
     model (light-blue, solid). The zero point is marked by a gray
     dashed line, dividing locations where convection (positive) or
     radiation (negative) is the dominant flux transport mechanism. 
     The inset zooms in on the deep interior where the magnetic field
     creates a small radiative core.}
    \label{fig:cmdra_del_dela}
\end{figure}

For fully convective stars, the situation is reversed. We display the run 
of $(\tgrad - \dela)$ in two models of CM Dra A in Figure \ref{fig:cmdra_del_dela}. 
One standard Dartmouth model and one magnetic model are shown. The magnetic 
model is a Gaussian radial profile model with a $6.0$ kG surface magnetic 
field. This is the same model that was plotted in Figure \ref{fig:cmdra_mag_d}(b). 
Fully convective stars are largely characterized by near-adiabatic convection
from the outer layers down to the core of the star. Changes in convective
properties have little effect on the flux transported by convection because
convection is extremely efficient. Changes to the properties of convection
do have some structural effects (see Figure \ref{fig:cmdra_mag_d}(a)), but they are minimal.
It's not until the deep interior magnetic field strength becomes strong enough
to stabilize interior regions of the star against convection ($\nu \sim 10^{-4}$) 
that significant structural changes occur (as with the Gaussian radial profile). 
When this occurs,
a large radiative shell appears in the interior, as shown by the inset in
Figure \ref{fig:cmdra_del_dela}. This radiative shell extends over 
54\% of the star by radius (between $0.18 R_{\star}$ and $0.72 R_{\star}$) and
78\% by mass (between $0.03 M_{\star}$ and $0.81 M_{\star}$).
We verified this was the dominant reason for structural changes by looking at 
the profile of a dipole radial profile model with a similar surface magnetic field 
strength. The $(\tgrad - \dela)$ profile exhibited in the surface layers by 
the dipole model is nearly identical to the Gaussian model. However,  
a radiative shell develops deep in the Gaussian model. This supports the idea that 
fully convective stars require radiative zones to be consistent with 
observations \citep{Cox1981,MM01}. 

A model with the constant $\Lambda$ profile has a slightly different
$(\tgrad - \dela)$ profile in the super-adiabatic layer. It also produces 
a marginally larger surface radiative zone than the standard
non-magnetic model. Convection in the deep interior is 
relatively unaffected, so a radiative core does not develop. We opted to 
not display these features in a figure because the overall profile is 
almost identical to the non-magnetic profile in Figure \ref{fig:cmdra_del_dela}. 
Given the insensitivity of the overall stellar structure of fully convective stars 
to the size of the super-adiabatic layer, the constant $\Lambda$ models 
have a negligible influence on the stellar radius.

\subsection{Surface Magnetic Field Strengths}
\label{sec:fc_sfs}

Surface magnetic field strengths are estimated from X-ray luminosity, $L_x$,
measurements using the relation between total unsigned magnetic flux, $\Phi$, 
and $L_x$  derived by \citet{FC13}. We are unable, however, to estimate a reliable 
magnetic field strength for Kepler-16\,B as it does not appear in the 
ROSAT catalogs. In a moment we will provide at least a reasonable upper bound,
but it is easiest to first address CM Dra.

The ROSAT Bright Source Catalogue \citep{Voges1999} indicates 
that CM Dra has an X-ray count rate of $X_{\rm cr} = 0.18\pm0.02\textrm{ cts s}^{-1}$ 
with a hardness ratio of ${\rm HR} = -0.30\pm0.07$. This translates into 
an X-ray luminosity per star of $L_x = (1.57\pm0.40)\times10^{28}\textrm{ erg s}^{-1}$, 
where we have used a parallax of $\pi = 68\pm4$\,mas \citep{Harrington1980}.
Note that the X-ray luminosities are upper limits due to possible X-ray 
contamination in the ROSAT data. There are several stars nearby to CM Dra 
in the plane of the sky, but it is difficult to judge whether they contribute
to the ROSAT count rate. 

From the X-ray luminosity derived above, we find $\log_{10}(\Phi/{\rm Mx}) = 24.81\pm0.45$.
Errors associated with the surface magnetic field strength estimates 
are substantial due to the large error on the surface magnetic flux. Converting
magnetic fluxes to surface magnetic field strengths, we estimate that 
$\langle Bf\rangle_{A} = 1.65^{+3.00}_{-1.07}$\,kG and
$\langle Bf\rangle_{B} = 1.85^{+3.36}_{-1.19}$\,kG for CM Dra A and B,
respectively. Note that quoted uncertainties are mean uncertainties. 
These estimates imply
that the $6.0$\,kG surface magnetic field strengths predicted by our Gaussian 
radial profile models are likely too strong. However, this does not invalidate
the magnetic field models, only our choice of radial profile. The magnetic
field strength in the deep interior is of greater consequence, so it may
be possible to construct a radial profile to greater reflect this fact.

Models that use a constant $\Lambda$ formalism predict surface magnetic
fields strengths up to $\sim3.0$ kG. This upper limit is set by the
magnetic field coming into equipartition with kinetic energy of convective 
flows \citep{Chabrier2006,Browning2008}. Values of $3.0$ kG are consistent 
with our X-ray estimated field strengths. Additionally, equipartition magnetic
field strengths are consistent with typical average magnetic field strengths 
measured at 
the surface of M-dwarfs \citep{Saar1996,Reiners2009,Shulyak2011,Reiners2012a}. 
Although magnetic field strengths are consistent with observations, our 
models are unable to produce radii consistent with these realistic field
strengths.

Although there are no X-ray measurements from ROSAT for Kepler-16, we
can attempt to derive a reasonable estimate. From photometry, we may 
estimate a distance to Kepler-16 by assuming that the primary contributes
to most of the observed flux in the visible. We estimate a distance of 
about 60\,pc using the temperature and luminosity provided by \citep{Doyle2011}
and \citet{Winn2011} in combination with a bolometric correction from
the {\sc phoenix} model atmospheres. If we make the assumption that 
all of flux at X-ray wavelengths is from the secondary \citep[recall the primary
shows only weak magnetic activity;][]{Winn2011}, then taking the ROSAT
sensitivity limit of $X_{\rm cr} = 0.005$ counts per second \citep{Voges1999}
we find $L_x \sim 2\times10^{28}$ erg s$^{-1}$. This also assumes ${\rm HR} = -0.1$,
typical for dwarf stars. Converting to an unsigned magnetic flux yields 
$\log_{10}(\Phi/{\rm Mx}) \sim 25$, or a magnetic field strength of order
2\,kG for Kepler-16\,B. Therefore, Kepler-16 likely does not have a magnetic
field strength any larger than that found on the surface of CM Dra. Furthermore,
if Kepler-16\,B is rotating pseudo-synchronously, like it's companion, then
it would have a rotation velocity $v\sin i < 0.5$\,km s$^{-1}$, as mentioned in 
Section \ref{sec:intro}. CM Dra, on the other hand, has $v\sin i \sim 9.0$\,km~s$^{-1}$.  
Kepler-16 may very well have a weaker magnetic field, owing to its longer orbital (and
presumably rotational) period. Therefore, we believe that the $6.0$\,kG 
surface magnetic field required to bring our models into agreement with 
observations is too strong.

\subsection{Comparison to Previous Studies}
\label{sec:comp_study}

Previous attempts to reconcile model radii with the observed radii of fully
convective stars have focused solely on CM Dra, as it was the only well-studied 
system known \citep{Chabrier2007,Morales2010,MM11}. In each case, magnetic 
fields and magnetic activity were found to provide an adequate solution,
which is quite the opposite conclusion from results presented in Section 
\ref{sec:ind_deb}. This seeming contradiction of previous results can be understood
by our neglect of star spots, in particular, the potential for observed
radii to be over-estimated on a spotted star. Before submitting that spots
are the solution and calling this case closed, we will review existing results,
placing ours into context, and the provide an assessment of the magnetic 
hypothesis.

\subsubsection{Summary of Methods \& Key Results}

Methods used in previous studies were, in some respects similar. Each used
a method for treating magneto-convection. Four techniques have been employed
thus far: a reduced-$\amlt$ approach \citep{Chabrier2007,Morales2010}, stabilization
of convection by a vertical magnetic field \citep{MM11}, stabilization by a
more general magnetic field (not specifically vertical; this work), and then
a turbulent dynamo approach that is similar to a reduced-$\amlt$ (this work).

Including effects of star spots is inherently difficult in a 1D stellar
evolution code. Spots are blemishes scattered across the stellar surface
that extend a non-fixed distance into the surface convection zone. Spots
have largely been treated in the same fashion in previous investigations
and rely on reducing the total stellar surface flux by a fractional amount,
\begin{equation}
    \beta = \frac{S_{\rm spot}}{S}\left[1 - \left(\frac{T_{\rm spot}}{T_{\rm phot}}\right)^4\right],
    \label{eq:spot}
\end{equation}
where $S_{\rm spot}/S$ is the surface areal coverage of spots and
$T_{\rm spot}/T_{\rm phot}$ is the spot temperature contrast. This approach 
is based on the ``thermal plug'' spot model advanced by \citet{Spruit1982a,
Spruit1982b} and \citet{Spruit1986}. The modified surface flux is then 
$\mathcal{F} = (1 - \beta)\mathcal{F}_\star$, where $\mathcal{F}_\star$ is the 
flux of the star if the photosphere is spot free. 

\citet{Morales2010} performed a detailed analysis to establish how specific
star spot properties effect results from both theoretical modeling and light
curve analyses. For a given star spot $\beta$ and assumed distribution of 
spots over the stellar surface (uniform, clustered at mid-latitudes, and 
clustered at the poles), they evaluated the reliability of light curve analyses
in determining stellar radii. If spots are preferentially located at the
poles, they showed that stellar radii may be over-estimated by as much as 
6\%. On the other hand, if spots are more evenly distributed across the 
surface, or clustered at mid-latitudes, radius determinations proved reliable
for $\beta < 0.3$. 

Combining the aforementioned results with the influence of spots on stellar 
evolution models, via Equation (\ref{eq:spot}), \citet{Morales2010} found that 
$\beta = 0.17$ was required to fit the stars of CM Dra. Magneto-convection 
using a reduced-$\amlt$ was found to be ineffective and was not required. Thus,
they predict that the stars in CM Dra are 35\% covered by spots that are 
15\% cooler than the surface (this latter value was fixed in their analysis).
This was deemed sufficient to correct model radii with observations.

\citet{MM11}, on the other hand, were able to produce agreement between their
model and observations using a combination of magneto-convection and star spots.
They identified regions of $\deltamm - \beta$ ($\beta\equiv f_s$ in their paper) 
parameter space that reconciles their models with CM Dra (see their Figures 
13 -- 15). Their magnetic inhibition parameter is defined as
\begin{equation}
    \deltamm = \frac{B^2}{B^2 + 4\pi\gamma P_{\rm gas}},
\end{equation}
where $B$ is the magnetic field strength, $\gamma$ is the ratio of specific
heats, and $P_{\rm gas}$ is the total gas pressure. This quantity is added 
directly to the adiabatic gradient in the Schwarzschild stability criterion. 
What strongly distinguishes the \citet{MM11} study from the \citet{Morales2010}
investigation is that \citet{MM11} do not include model inflation due to 
spots as described by Equation (\ref{eq:spot}). Instead, they only adopt
the adjustment made to the observed stellar radii due to the influence of
polar spots on the light curve analysis. Model radius inflation is then caused
by stabilization of convection using their inhibition parameter described
above.

The full range of values for which they find agreement is 
$0.15\lesssim\beta\lesssim0.28$ with $0.0 < \deltamm < 0.025$. To further
constrain the parameter space, they assume the best fit star spot $\beta = 0.17$
from \citet{Morales2010}. This narrows the acceptable range for the magnetic
inhibition parameter to $0.020 < \deltamm < 0.025$. These values of $\deltamm$ 
correspond to vertical magnetic field strengths of approximately 500\,G
at the stellar photosphere and a (capped) interior vertical magnetic field
strength of 1\,MG. 

Our modeling results are largely consistent with those of \citet{Morales2010}
and \citet{MM11}. Comparing first with \citet{MM11}, we focus on the dipole
and Gaussian radial profile models with a rotational dynamo (Figures 
\ref{fig:cmdra_mag_d}(a) and (b)). The magnetic perturbation applied in 
these models is qualitatively similar to magneto-convection method favored 
by \citet{MM11}. We have estimated that our magnetic field strengths should 
be up to an order of magnitude larger than those of \citet{MM01} as a 
consequence of our formulation \citep{FC13}.
Our ``inhibition parameter'' is primarily controlled by the quantity,
\begin{equation}
	\nu\delx 
		= \frac{P_{\rm mag}}{P_{\rm gas} + P_{\rm mag}}\left(\frac{d\ln\chi}{d\ln P}\right),
\end{equation}
where $\nu$ is a magnetic compression coefficient and $\delx$ is the gradient
of the magnetic energy per unit mass with respect to the total pressure. We 
showed that $\nu \sim \deltamm$ \citep{FC13}, but
our formulation has the additional term  $\delx \sim 0.1$. Therefore, 
to achieve the same results as \citet{MM11}, we should need $\nu$ to be 
about 10 times larger than $\deltamm$. Thus, our requirement of a 6.0 kG
surface magnetic field strength with a roughly 40 MG peak magnetic field
strength for the Gaussian radial profile is consistent with the values 
we would expect given the values from \citet{MM11}.
Although the dipole radial profile model has a 5.0 kG surface magnetic 
field strength, the peak magnetic field strength is only 1.5 MG, a factor 
of 10 or so too small to impart the necessary structural changes.

To compare with \citet{Morales2010}, we refer to our constant $\Lambda$ 
models, which closely matches the reduced-$\amlt$ formulation. We found 
that even if the magnetic field were a considerable fraction
of the equipartition magnetic field strength ($\Lambda = 0.9999$), there
is little impact on the stellar radii. This is consistent with \citep{Morales2010},
who found that the reducing $\amlt$ has a negligible impact on the stellar
model predictions for fully convective stars, especially after star spots 
were invoked. In that sense, we confirm that reducing convective efficiency 
does not appear to be a viable method to fully account for inflation among 
fully convective stars.

Note that we have not adopted the reduced stellar radius presented by 
\citet{Morales2010} nor have we explicitly addressed reductions in surface 
flux due to star spots. Models presented in \citet{FC13} did not require star 
spots to provide an adequate fit to the data. This reinforces a known degeneracy 
between magneto-convection implementations and introducing star spots \citep{MM10}. 
One may not be too surprised that these two issues are so closely connected, as
spots are the physical manifestation of suppressed convection near the stellar
surface. Therefore, it is reasonable to assume that reduction in convective flux
from either magneto-convection or star spots should produce similar results.
The degeneracy, in a sense, can be broken by the idea that star spots could 
bias light curve analyses towards larger radii. Magneto-convection, strictly
speaking, has no influence on the light curve analysis, if we assume global 
changes to stellar properties. In that sense, particular configurations of 
spots (clustered near the pole) could have a significantly larger impact on
the mass-radius problem than magneto-convection. 

Despite the aforementioned results that support the magnetic field hypothesis, 
we are skeptical of the interpretation that magnetic fields are driving the 
observed radius inflation. We will now assess the results primarily by considering 
available observational data.

\subsubsection{Assessment of Magneto-convection}
\label{sec:mag_conv}

Reductions in convective efficiency appear to be inadequate for producing 
the observed stellar radii \citep[Section \ref{sec:ind_deb} of this work;
][]{Chabrier2007,Morales2010}. We showed that, for Kepler-16 and CM Dra,
radius inflation induced by the inhibition of convective efficiency was 
effectively negligible ($\sim$0.1\%). This is in agreement with \citet{Chabrier2007}
and \citet{Morales2010}, who find reducing the convective mixing length
from $\amlt = 2.0$ to $\amlt = 0.1$ has little effect on the radii
of fully convective stars. Reducing $\amlt$ further could begin to provide 
reasonable effects on fully convective stellar radii. The question then
becomes, ``is it possible for $\amlt \rightarrow 0$, and if so, by what
mechanism?''

Results from the stabilization of convection are encouraging. A proper 
amount of radius inflation can be achieved with somewhat reasonable surface 
magnetic field strengths \citep[Section \ref{sec:ind_deb} of this work;][]{MM11}. 
Instead of surface field strengths posing a problem as they did with 
partially convective stars \citep{FC13}, it is the interior magnetic 
field strengths, which range from 1~MG to 50~MG, that require attention.

There is presently no observational evidence to suggest a 1 MG (or greater) 
magnetic field could exist within a fully convective star. To be fair, 
there is also no direct evidence to rule out the possibility. We must be honest
about the lack of observational data concerning interior magnetic field 
strengths. However, there is some indirect evidence that casts doubt on
the existence of super-MG interior magnetic fields. We will now carefully 
examine the possibility that such fields do exist deep in fully convective
stars.

From a theoretical perspective, there is concern about how a super-MG
magnetic field is generated. Simulations suggest that turbulent dynamos
reach saturation at field strengths of $\sim$50\,kG \citep{Chabrier2006,
Dobler2006,Chabrier2007,Browning2008}, two orders of magnitude below
the 1\,MG magnetic field required by stellar models. We will return to
this a bit later. For now, let us assume that super-MG magnetic fields
cannot be generated by dynamo action. Instead, it could be assumed that
super-MG magnetic fields are the result of the amplification of a primordial
$\mu$G seed magnetic field during the proto-stellar collapse.

Assuming magnetic flux is conserved during collapse of the proto-star, 
one finds that super-MG magnetic fields could plausibly exist. However,
the super-MG magnetic fields must then survive several Gyr within the
stellar interior without decaying. There are three primary timescales 
of interest: that given by macroscopic diffusion in a non-convecting 
medium, the rise time for a buoyantly unstable flux tube, and
the advection timescale in a convecting medium with relatively 
high conductivity. We'll consider the latter, first. 

If the stellar plasma is highly conducting, which is a valid assumption
throughout the interior of stars, then the field lines will be frozen into 
the plasma. We might then expect that the flux tubes will be carried
by convection from deep within the star to the stellar surface, where the
they will then dissipate their energy. The timescale for this to occur
is roughly $\tau_{\rm conv} = R_{\star}/\uconv$, where $R_{\star} \sim 10^{10}$
cm and $\uconv \sim 10^{3}$ cm s$^{-1}$. We find that $\tau_{\rm conv} \sim 10^{7}$ s,
or about 1 yr. Even if we assume $\uconv \sim 10^1 - 10^2$ cm s$^{-1}$, 
$\tau_{\rm conv} \sim 10^{8} - 10^{9}$ s. Thus, the timescale for this 
type of advection is about 10 yr. Although convection can quickly carry
a flux tube from the deep interior to the stellar surface, the process 
can also carry flux tubes from the surface to deep in the interior.

A more uni-directional process that can transport magnetic flux to the 
stellar surface is the magnetic buoyancy instability \citep[e.g.,][]{Parker1955,
Parker1979}. 
Magnetic flux tubes are assumed to be in pressure equilibrium with their 
surroundings deep in the interiors of stars,
\begin{equation}
	p_{{\rm gas},\, i} + \frac{B^2}{8\pi} = p_{{\rm gas},\,e},
\end{equation} 
where $p_{{\rm gas},\, i}$ and $p_{{\rm gas},\,e}$ are the interior and
exterior gas pressures acting on the flux tube, respectively, and 
$B^2/8\pi$ is the magnetic pressure within the flux tube. Since 
flux tubes are supported by both
the internal gas pressure and the magnetic pressure, the gas density 
within a flux tube is lower than the surrounding gas pressure, provided
that the temperature inside the flux tube is the same as the temperature
of the surroundings. This density perturbation creates a buoyancy force
towards the surface of the star, but will be counteracted by radially 
inward hydrodynamic forces resulting from convective down-flows. 

If we assume a polytropic EOS, a reasonable approximation for
deep stellar interiors, and that $p_{{\rm gas}}$ can be expressed as a 
power-series expansion about $\rho_{{\rm gas},\,e}$, the gas density exterior
to the flux tube, then truncating the series to first order gives
\begin{equation}
	p_{{\rm gas},\,e}\left[1 + \gamma\rho_{{\rm gas,\, e}}^{-1}\Delta\right]
		+ \frac{B^2}{8\pi} = p_{{\rm gas},\,e},
\end{equation}
where $\gamma$ is the ratio of specific heats and 
$\Delta = \rho_{{\rm gas,\, i}} - \rho_{{\rm gas,\, e}}$ is the density 
perturbation. The specific buoyancy force resulting from this density 
perturbation is 
\begin{equation}
	f_b = \frac{g}{\gamma}\left(\frac{B^2}{8\pi p_{{\rm gas},\,e}}\right).
\end{equation}
Balancing the buoyancy force with the hydrodynamic force \citep{Fan2009}
yields the condition whereby a magnetic flux tube will be unstable to
buoyant rise,
\begin{equation}
	B \gtrsim \left(\frac{H_P}{a}\right)^{1/2} B_{\rm eq},
\end{equation}
where $B$ is the magnetic field, $H_P$ is the local pressure scale height, 
$a$ is the flux tube radius, and $B_{\rm eq}$ is equipartition magnetic field 
given in Section~\ref{sec:clambda}. This is a simplified picture
that ignores effects due to curvature and the response of the magnetic tension
that may act to prevent the onset of buoyant instability. However, it provides
a reasonable order of magnitude approximation. Given the propensity for magnetic stellar
models to invoke magnetic fields with $B \ge 10^6$~G, we can estimate a typical
flux tube radius needed to maintain this magnetic field strength stable against
buoyancy. 

For M-dwarfs, $H_P \sim 10^9$~cm and $B_{\rm eq} \sim 10^4$~G. 
Thus, to maintain strong $10^6$~G magnetic fields deep in the interior 
\citep[as in][]{MM11}, flux
tubes must be no larger than $10^5$~cm, or 1~km. This means the ratio of the
pressure scale height to the flux tube radius must be $\sim 10^{4}$. By 
comparison, the values estimated for this ratio near the solar tachocline
are of order unity \citep{Fan2009}. This ratio rises to $10^6$ ($a \sim 10$~m)
if the deep interior magnetic field is to be of order $10^7$~G (this work).
Once the flux tube becomes unstable to buoyant rise, an estimate of the 
rise time can be approximated assuming that the tube travels at the 
Alfv\'{e}n velocity \citep{Parker1975}. A $10^6$~G magnetic field will 
traverse a stellar radius $R \sim 10^{10}$~cm in approximately $10^{5.5}$~s, 
or about 10 days, assuming a constant density of $10^{2}$~g~cm$^{-3}$. 
This rise time will be increased by non-adiabatic heating effects,
but remains small compared to evolutionary timescales \citep{Parker1974}.

Figure \ref{fig:cmdra_del_dela} indicates that a radiative shell
develops deep within the star in the presence of a super-MG magnetic field. This
means that the dissipation time of the magnetic field does not necessarily
obey the advection timescale discussed above. Instead, we may look at the
timescale for diffusion given by the induction equation. In the absence of
any favorable current networks, we have that
\begin{equation}
    \frac{\partial\bf B}{\partial t} = \eta{\bf \nabla}^2{\bf B},
\end{equation}
where $\eta$ is the magnetic diffusivity. The approximate timescale for 
diffusion, assuming $\eta \sim 10^2 - 10^3$ cm$^2$ s$^{-1}$ \citep{Chabrier2007}, 
is then $\tau_{\rm diff} \sim L^2/\eta$, where $L$ is the size of the radiative
shell. The field diffuses through the radiative shell until it reaches the
convective boundary, which then efficiently transports magnetic flux to 
the stellar surface. Based on our models of CM Dra, we find the size of
radiative shell is $L \sim 10^{10}$ cm. Therefore, 
$\tau_{\rm diff} \sim 10^{10} - 10^{11}$~yr, assuming the magnetic field 
fully diffuses out of the radiative zone. However, even within a radiation
zone, magnetic buoyancy must be considered and can lead to a rapid rise
time for magnetic flux tubes \citep[e.g.,][]{MacGregor2003}.

These are only order of magnitude approximations, so it is possible that
the field could survive for a shorter or longer time in any scenario. 
Instead, this exercise suggests that an amplified seed field, or any 
super-MG magnetic field, present in a fully convective star will likely 
decay away rapidly owing to the magnetic buoyancy instability. We note 
that this is still possible, even if a sizeable radiative zone develops 
as a result of the strong magnetic field. Furthermore, the radiative zone 
is a product of the magnetic field and therefore exists in an unstable 
equilibrium. As the magnetic field decays away, so will the radiative zone.

\citet{MM11} recognized that primordial magnetic fields would likely
decay and thus were not a suitable solution for the existence of super-MG
magnetic fields within CM Dra. Instead, they proposed that the dynamo
mechanism may be able to generate the necessary magnetic fields. Their
proposal contradicts what we mentioned earlier, that simulations are 
unable to generate magnetic fields greater than $\sim$50\,kG in fully
convective models. \citet{MM11} reason that this is a result of those
simulations using rotational velocities similar to those in the Sun.
If one nominally assumes that magnetic field strengths increase with
rotation rate, then a star like CM Dra will have considerably stronger
magnetic fields than does the Sun. 

Using a scaling relation between magnetic field strength and rotational
angular velocity, \citet{MM11} find that CM Dra should have a surface
magnetic field strength of about 500\,G with an internal magnetic field
strength anywhere between 0.3\,MG and 1.3\,MG. This could then explain
the existence of super-MG magnetic field strengths. Note that the
magnetic field strength is predicted to be the large-scale magnetic field
strength, not necessarily the total magnetic field strength.

However, the peak interior
magnetic field strength seen in simulations is moderated by the kinetic 
energy available in convective flows. This was discussed in Section
\ref{sec:clambda}. Simulations find that a dynamo driven by turbulent 
convection can create both small-scale magnetic fields and large-scale
magnetic fields \citep{Chabrier2006,Browning2008}. It is not immediately
clear that rotation would not drive stronger magnetic field strengths,
so we will proceed in our discussion assuming that it might to see if
we encounter any inherent contradictions.

The scaling relation adopted by \citet{MM11} would then also explain
why KOI-126 can be fit by stellar evolution models. KOI-126 is a triply
eclipsing system found earlier in the \emph{Kepler} data \citep{Carter2011}
that contains two fully convective stars that are nearly identical to 
CM Dra. Yet, the properties of KOI-126 (B, C) are well reproduced by
standard stellar evolution models \citep{Feiden2011,SD12}. Assuming a
scaling relation between magnetic field strength and rotational angular 
velocity may support this finding. KOI-126 (B, C) have an orbital period
of 1.77\,days, to be compared to the 1.27\,day orbit of CM Dra. This 
0.5\,day difference in orbital period means KOI-126 (B, C) would have 
weaker magnetic fields than CM Dra and would experience considerably
less radius inflation \citep{MM11}.

\citet{MM11} predict that KOI-126 should be inflated by approximately 
2 -- 3\%. We find this difficult to understand in context of the
agreement between standard stellar evolution models and the observed
radii. Inflating model radii by 2 -- 3\% would mean that models would
over-predict the radii of KOI-126 (B, C). If one assumes that spots are 
biasing the observed radius measurements, it would decrease the measured 
radii, again breaking the agreement between the observations and the 
models. There seems to be problems with simultaneously increasing model
radii and decreasing the measured radii of KOI-126 (B, C) while still 
maintaining model-observation agreement. 

Furthermore, the scaling relation suggested to validate super-MG magnetic
fields may not be compatible with observations. Results seem to show that
at a given mass, the large-scale magnetic field strength of fully convective 
stars do not appear to depend on rotation below a critical threshold 
\citep[see, e.g., Figure 3 in][]{Donati2009}. For low-mass
stars, this threshold corresponds to a rotation period of about 3 or 4
days. Magnetic field saturation is also observed in data regarding
the total magnetic field strength of fully convective stars. Magnetic fields 
saturate around 3\,kG for $v\sin i > 3$\,km\,s$^{-1}$ \citep{Reiners2009a,
Shulyak2011}. Saturation inherently
implies that scaling relations are invalid. Since both CM Dra and KOI-126
(B, C) have components rotating faster than the threshold for saturation,
there is no reason to assume that their magnetic fields are that different.
There is the possibility that their large-scale components (and thus vertical)
magnetic fields are sufficiently different, but this is not immediately 
obvious. 

Finally, we note that magneto-convection techniques alone have not been
sufficient to rectify the model-observations disagreements. Effects of
star spots must be included and must therefore be assessed on their own
basis.

\subsubsection{Assessment of Star Spots}
\label{sec:spots}
The most efficient means of inflating low-mass stellar models of CM Dra 
is by including star spots \citep{Morales2010,MM11}. Although we did not
explicitly include spots, we will take a careful look at the results of these 
two previous studies. We feel this analysis is required in light of recent 
observational data, including confirmation that CM Dra has a sub-solar 
metallicity of [Fe/H] $\approx -0.3\pm0.1$ \citep{RojasAyala2012,Terrien2012}.

We use approximate fits to the \citet{Morales2010} spot-modeling results
to estimate how a metallicity reduction alters their conclusions. Assuming
polar spots, they find that spots tend to bias observations towards larger 
radii. From the few data points in their Figure 7, we calculate that spots
produce larger radii according to the function
\begin{equation}
	R_{\rm spot}/R_{\rm real} = 0.187\beta + 1.0,
    \label{eq:rad_bias}
\end{equation}
where $R_{\rm spot}/R_{\rm real}$ is the ratio of the radius derived when 
polar spots are present to the actual radius. This function can be used to 
correct for the radius bias produced by spots. In stellar models, spots 
can also increase radius predictions. Given the data in Figure 8 of 
\citet{Morales2010}, model radii can be approximated as,
\begin{equation}
	R/\rsun = R_0\left(0.172\beta^2 + 0.0834\beta + 1.0\right)
	\label{eq:rad_inf}
\end{equation}
where $R_0$ is the radius of a non-magnetic model in units of solar radii. We 
note that this formula approximates the effects of spots only for model masses 
of $\sim0.23\msun$. The intersection of these two relations in the $\beta$--$R$
plane provides an estimate of the star spot parameter required to correct models.

\begin{figure}
	\plotone{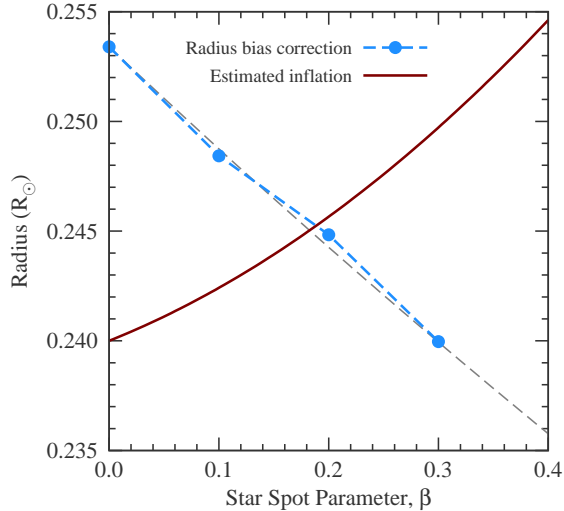}
	\caption{Effects of star spots on DEB radius measurements and on stellar
	         evolution model radii. Data points in light-blue, connected by
	         a dashed line, show the downward corrected radius for CM Dra A in 
	         the presence of polar spots. Data are taken directly from Figure~7 
	         in \citet{Morales2010} for a polar spot distribution. The light-grey 
	         dashed line is a linear fit to the data given in Equation 
	         (\ref{eq:rad_bias}). The solid maroon curve
	         illustrates how model radii increase with spot coverage.}
	\label{fig:spot_corr}
\end{figure}

Figure \ref{fig:spot_corr} illustrates that a model of CM Dra A with [Fe/H] 
$= -0.3$ (see also Figure~\ref{fig:cmdra_nmag}) needs a star spot 
$\beta \sim 0.19$ to match observations. Small differences in our results from
those of \citet{Morales2010} are due to the adopted base model (Dartmouth vs
Lyon) and our method of fitting polynomials to the spot radius bias and inflation
data in \citet{Morales2010}. Changing $\beta$ has proportionally larger effect 
on correction for radius measurement biases than it does on inflation caused 
in stellar models. If this correction is not applied, $\beta = 0.38$ is needed 
to produce agreement using model radius inflation at sub-solar metallicity.

Whether the required star spot parameter is realistic is a more difficult 
question. Spot properties for M-dwarfs are relatively unconstrained, as 
identifying spot sensitive features in spectra is complicated by uncertainties
in modeling molecular features. Nonetheless, there are three components to 
the spot parameter that are of interest:
spot coverage, spot temperature contrast, and spatial distribution. We will 
address each separately and then present a unified picture afterward.

\paragraph{Temperature Contrast}
Spot temperature contrast is defined, here, to be the ratio of the spot 
temperature to the temperature of the unblemished stellar photosphere. As 
seen from Equation (\ref{eq:spot}), constraining the spot temperature 
contrast provides insight into the required filling factor predicted by the 
spot model. Although no direct empirical constraints exist for M-dwarfs, 
there have been studies performed to measure spot temperature contrasts 
of K stars using TiO bandhead features \citep{ONeal1998,ONeal2004}. We 
will take a closer look at those results, here.

Properties of spots were determined by fitting the shape of TiO bandheads
with a spectrum created using a two temperature spot model. This technique
involves convolving two template spectra for which the effective temperature
is assumed to be known and finding the best combination of template spectra
to produce the molecular bandhead features in spectra of a given star
\citep[see e.g.,][]{ONeal1998}.
The result is an estimate of the temperatures for the unblemished photosphere
and the spotted regions, as well as the fractional surface coverage of each
feature (see below). Using this technique, it was found that active K 
stars---specifically II Peg and EQ Vir---have spots with a characteristic 
temperature contrast of 75\% (25\% cooler than the unblemished photosphere). 

This estimate, however, relies on the temperature for the two template spectra
being correct. Results from interferometric studies \citep[e.g.,][]{Boyajian2012} 
have revised the 
temperature estimates for the stars whose spectra were used as templates in 
the \citeauthor{ONeal1998} investigations. These revisions only affect the 
temperatures for template spectra used as the spot (cool) component in the fit, namely 
the temperatures associated with cooler M-dwarf spectra, and increase the
associated effective temperature by 300~K -- 400~K. Taking into account the
revised temperature scale for the cooler template spectra, one finds the
temperature contrast for K stars inferred from molecular bandhead features
is between 82\% (II Peg) and 90\% (EK Vir). The warmer spot temperatures
are supported by Doppler Imaging (DI) studies of II Peg, which indicate temperature
contrasts between 80\% and 85\% \citep{Hackman2012}.

Assuming that spot temperature contrasts become weaker towards later spectra
types \citep{Berdyugina2005}, then one might expect M-dwarfs to have spot 
temperature contrasts around 90\% (10\% cooler than the unspotted surface).
Using this value, one finds that a spot surface coverage of 55\% is required
to satisfy the condition that $\beta = 0.19$ for CM Dra. Of course, slightly 
warmer or cooler spots may be permitted, depending on the efficiency at which
spots temporarily suppress surface convection. For the sake of argument, we 
take 90\% to be a typical spot contrast. We now look whether a spot areal 
coverage of 55\% is consistent with empirical evidence. 

\paragraph{Surface Coverage}
To assess what constitutes typical spot surface coverages, we look to results
from DI and molecular bandhead fitting. Using DI, spot
coverages for three M-dwarfs have been estimated. DI maps of \object{HK Aqr}, 
\object{RE 1816+541}, and \object{V374 Peg} \citep{Barnes2001,Barnes2004,
Morin2008v374} reveal varying levels of surface coverage from 2\% (V374 Peg) 
up to 40\% (HK Aqr).
Similarly, surface coverages derived by modeling the shape of TiO bandheads 
of active K stars---again II Peg and EK Vir---are typically around 
30 -- 40\%, consistent with
values found in DI investigations \citep{ONeal1998,ONeal2004}. There are no 
measurements of spot coverages on M-dwarfs using this technique, which 
limits the applicability of these results, but we may cautiously 
extrapolate and assume that M-dwarfs are able to possess at least 
similar surface coverages. 

For one of the K stars studied using the molecular bandhead technique, 
\object{II Peg}, information on spot coverage from DI is also available.
\citet{Hackman2012} mapped the spot coverage of II Peg and find seasonal
variation in spot coverage between 5 -- 20\%, a factor of 8 and 2 lower 
than the coverage inferred from TiO bandheads, respectively. Seasonal 
variation is expected if magnetic activity cycles are present in other 
stars, so it's not surprising that one observes this phenomenon. However, 
the difference between the maximum surface coverage revealed
by DI (20\%) and that obtained using molecular bandhead features (40\%) 
highlights a known limitation with DI studies: they are only sensitive to 
spot features larger than the grid resolution in the DI analysis. In other 
words, small scale spot features will be missed by the DI reconstruction. 
This is especially relevant since comparison of spectro-polarization 
measurements of M-dwarf magnetic fields using Stokes $I$ and $V$ polarization
indicate a majority of the magnetic energy is contained in small-scale 
features \citep{Reiners2009}.
Therefore, spot coverages inferred from DI maps must be taken as lower 
limits to the true spot coverage. One may then plausibly expect spot 
surface coverages at least as high as 40\% for M-dwarfs.

Evidence from both DI and molecular bandhead fitting support
the existence of spot coverages around 40\% for active late K-dwarfs and
M-dwarfs. This is below the 55\% needed to satisfy the use of a $\beta = 0.19$ 
spot parameter. Whether 55\% is {\it unrealistically} high is unclear. If, for instance,
the spot temperature contrast were around 85\%, as opposed to 90\%, then the
required spot surface coverage would be 40\%, and thus consistent with observations.
One could just as well argue the opposite, that a larger temperature contrast
of 94\% would require a surface coverage of 87\% and push the requirement further
from the current empirical evidence. 

One further scenario is that stars are able to achieve spot coverages near
100\%. Observations suggest that the filling factor of magnetic fields at the 
surface of fully convective stars (both main-sequence and pre-main-sequence) 
approaches unity \citep{JohnsKrull2004,Shulyak2011,Johnstone2014}. We must then 
try to understand the connection between the presence of magnetic fields and 
star spots. This includes determining how strong of a magnetic field is necessary 
at the surface of an M-dwarf to generate spots of appreciable temperature contrast.
It may be that pervasive 1~kG magnetic fields covering the whole surface of 
fully convective stars may be too weak to cause any noticeable effects on 
convection. Such an even distribution of small spots would also likely not 
cause any noticeable light curve modulation, if filling factors approach unity.
However, pockets of strong 5 -- 8~kG magnetic fields \citep{Shulyak2011} may
produce the noticeable light curve modulation, bias stellar radii measurements,
and suppress surface convection.

\paragraph{Spatial Distribution}
Arguably, whether spots cluster near the poles of active M-dwarfs is of 
greatest consequence for the present assessment.
Evidence for spots clustered at the poles is less definitive than temperature
contrasts and surface coverages. About half of all stars studied using DI 
show signs of polar caps \citep{Berdyugina2005}. 
Exceptions are very active stars and M-dwarfs, which tend to have
spots distributed across all latitudes. Of the three DI spot maps generated
for M-dwarfs, none display evidence for polar caps \citep{Barnes2001,Barnes2004,
Morin2008v374}. A DI study of CM Dra to search for possible signs of 
significant polar coverage of spots is needed. Ideally, this
would be performed with existing data that was used to derive the stellar 
properties, but a look at the signal-to-noise of the observations reveals that
the data is likely unsuitable for such an investigation \citep[see][]{Metcalfe1996}. 
In the meantime, we must be cautious as even polar spots would require large 
areal coverages (half or more of the stellar surface) to provide agreement 
between models and observations. 

Additional evidence for non-polar-cap distributions comes from studies of
the $\sim$150 Myr open cluster \object{NGC 2516}. The statistical distribution 
of light curve modulations assumed to be caused by star spots in the 
low-mass star population can only be reproduced if the stars have randomized 
spot latitudinal distributions and rotational inclination angles 
\citep{Jackson2009,Jackson2013}. We do note that these latter studies also 
require spot temperature contrasts of 75\%, which we have just discussed 
may be too dramatic for K- and M-dwarfs. How using a more realistic spot 
contrast would alter their results is unclear as there are strong degeneracies 
when modeling spot modulation in light curves. A first estimate would be 
that typical surface coverages would increase from the $40\pm10$\% that
was required in their study. 

There is as yet no definitive empirical evidence for polar cap spots
among fully convective main sequence stars. However, there is a 
theoretical expectation that spots should be located near the poles
in rapidly rotating stars \citep[e.g.,][]{Schuessler1992,DeLuca1997}. 
Whether this expectation applies to fully convective stars is questionable,
as spots distributed more evenly at lower latitudes may be more reasonable
for more distributed dynamos \citep{DeLuca1997}. Based on only three data 
points, the apparent lack of polar spots among fully convective main sequence 
stars is not robust and requires further confirmation.

Is it then possible to abandon the idea of a largely polar cap distribution 
and still rely on spots to address the radius discrepancies? In Figure 
\ref{fig:spot_corr}, we see that abandoning the idea of polar spots 
requires $\beta = 0.38$ if model inflation is to alone correct for the 
observed radii. In this case, a spot temperature contrast of 89\%
with areal coverage of 100\% would be required. Temperature contrasts
greater than 89\% lead to areal coverages greater than 100\% and are 
thus unphysical. Even as one approaches 100\% surface coverage (for 
temperature contrasts below 89\%), it becomes increasingly difficult
to produce spot modulation in light curves as there are fewer regions
of unblemished photosphere left to produce the asymmetries needed, 
as discussed in the previous section. 
Considering patches of varying spot temperature contrast may provide
the necessary asymmetries, but has not yet been investigated for 
consistency with the empirical data. 

\subsubsection{Putting it all Together}
The case for magnetic fields inflating the radii of fully convective stars
appears to be tenuous. Stabilization of convection in stellar models can 
produce the necessary radius inflation with reasonable surface magnetic 
field strengths. But, the models require super-MG magnetic fields in the 
interior. We are hesitant to suggest that real stars contain such strong 
interior magnetic fields for several reasons. 

First, turbulent dynamos cannot generate magnetic fields with 
strengths greater than about 50\,kG \citep{Dobler2006,Chabrier2006,Browning2008}.
How magnetic fields with strengths greater than 1~MG would accumulate in 
the interior of fully convective stars would be unknown. If super-MG magnetic 
fields are of primordial origin, they would probably not survive to the 
present day. Nor would magnetic field strengths of that magnitude be generated 
within the star and survive on evolutionary timescales. To avoid this, one
can assume a scaling relation between magnetic field strengths and rotation, 
as suggested by \citet{MM11} to explain the super-MG magnetic fields. However,
this breaks down for most stars in DEBs. Observations indicate a saturation of 
magnetic flux for fully convective stars that have rotation periods below 
$\sim2.5$\,days \citep{Donati2009,Reiners2009a,Shulyak2011}. Also, KOI-126 (B, C) 
agree with stellar evolution models \citep{Feiden2011}. These stars exist in 
a regime where magnetic flux saturation should occur, but are seemingly uninfluenced 
by the presence of any magnetic phenomena. If super-MG magnetic fields are causing 
dramatic inflation in CM Dra, one would expect to observe it in KOI-126 (B, C) 
as well.
Finally, suppression of convective efficiency is unable to provide a suitable 
solution on its own. Instead, star spots may be invoked for fully convective 
stars, which introduces additional concerns.

The presence of star spots on the stellar surfaces may provide an adequate 
solution, mostly through biasing the analysis of light curve data. From 
observations, star spots on fully convective stars can be characterized
as: having temperatures about 10\% cooler than the surrounding photosphere
\citep{ONeal1998,ONeal2004,Berdyugina2005,Hackman2012}, having surface coverages 
ranging from a couple percent \citep{Morin2008v374,Hackman2012} up to around 40\%
\citep{Barnes2001,Barnes2004,ONeal1998,ONeal2004}, and having spot distributions 
that appear to be more random and not clustered at the poles.

At this point, there is not sufficient empirical data regarding spot properties
of M-dwarfs to draw firm conclusions. Still, the lack of noticeable spots at the 
poles is a concern. Results regarding how spots bias radius measurements rely on 
concentrated regions of spot coverage at the poles that are darker than their 
surroundings. At present, this feature is not observationally supported, but more 
work needs to be performed on this matter.

\section{Summary}
\label{sec:summ}

This paper addressed the question of whether magnetic fields can realistically
inflate the radii of fully convective stars. We approached the problem by
modeling two individual DEBs that contain fully convective stars, Kepler-16
and CM Dra. Our models showed that the observed radii could be reproduced
provided that a strong 50~MG magnetic field resides deep within these
fully convective stars. We proceeded to show that multiple mechanisms---advection 
due to convective flows, macroscopic diffusion, and magnetic buoyancy 
instability---can lead to rapid destruction of such strong magnetic fields. 
These estimates would benefit from detailed numerical simulations to probe
further whether such magnetic fields can stably reside deep within M-dwarfs.
An assessment of the hypothesis that star spots are responsible for 
the observed inflation was also given. Here, the hypothesis largely rests on
the assumption that large conglomerations of spots lie near the poles;
a prediction that has not yet been observed among main sequence fully
convective stars. 

After weighing the evidence, we find it difficult to favor the hypothesis
that magnetic fields are actively inflating the radii of fully convective,
main sequence stars. There are numerous uncertainties that must be addressed
on both the theoretical and observational side of the hypothesis. 
More sophisticated magneto-convection models are required to further investigate
this issue, using both simplified 1D approximations and more detailed 3D MHD 
models. Most critical, though, will be observational studies of star spots
and magnetic fields of low-mass stars. These properties must be further 
constrained by observations if we are to rule out any of the physical pictures
presented by the current generation of 1D stellar evolution models. In the
meantime, it may not be too early to start seeking other solutions.

\acknowledgements

We thank Bengt Gustafsson for carefully reading and commenting 
on the manuscript, and the anonymous referee for their constructive 
remarks and suggestions.
This research was made possible by the William H. Neukom 1964 Institute 
for Computational Science at Dartmouth College and the National Science 
Foundation (NSF) grant AST-0908345. This work made use of NASA's Astrophysics 
Data System, the SIMBAD database, operated at CDS, Strasbourg, France, and 
the \emph{ROSAT} data archive tools hosted by the High Energy Astrophysics 
Science Archive Research Center (HEASARC) at NASA's Goddard Space Flight 
Center.

\bibliographystyle{apj}

\end{document}